\documentclass[preprint,eqsecnum,preprintnumbers,nofootinbib,byrevtex,prd,aps,showpacs,showkeys,groupedaddress,floatfix]{revtex4}
\usepackage{bm}
\usepackage{graphics}
\usepackage{graphicx}
\usepackage{epsfig}
\usepackage{amssymb}
\usepackage{amsmath}
\begin{document}
\title{THE MESON PRODUCTION IN PROTON-PROTON COLLISIONS IN NEXT-TO-LEADING ORDER AND INFRARED RENORMALONS}
\author{A.~I.~Ahmadov$^{1,2}$~\footnote{ahmadovazar@yahoo.com}}
\author{R.~M.~Burjaliyev$^{2}$}
\affiliation{$^{1}$
The Abdus Salam International Centre for Theoretical Physics\\
Strada Costiera 11, 34014, Trieste, Italy}
\affiliation{$^{2}$ Department of Theoretical Physics, Baku State
University  \\ Z. Khalilov Street 23, AZ-1148, Baku, Azerbaijan}
\begin{abstract} In this article, we investigate the next-to-leading order
contribution of the higher-twist Feynman diagrams to the
large-$p_T$  inclusive pion production cross section in
proton-proton collisions and present the general formulae for the
higher-twist differential cross sections in the case of the
running coupling and frozen coupling approaches. We compared the
resummed next-to-leading order higher-twist cross sections with
the ones obtained in the framework of the frozen coupling approach
and leading-twist cross section. The structure of infrared
renormalon singularities of the higher twist subprocess cross
section and it's resummed expression (the Borel sum) are found. It
is shown that the resummed result depends on the choice of the
meson wave functions used in the calculations. We discuss the
phenomenological consequences of possible higher-twist
contributions to the meson production in proton-proton collisions
in next-to-leading order at RHIC.
\end{abstract}
\pacs{12.38.-t, 13.60.Le, 13.87.Fh, 14.40.Aq,}

\keywords{higher-twist, pion wave function, infrared renormalons}
\maketitle

\section{\bf Introduction}
The large-order behavior of a perturbative expansion in gauge
theories is inevitably dominated by the factorial growth of
renormalon diagrams [1-4]. In the case of quantum chromodynamics
(QCD), the coefficients of perturbative expansions in the QCD
coupling $\alpha_{s}$ can increase dramatically even at low orders.
This fact, together with the apparent freedom in the choice of
renormalization scheme and renormalization scales, limits the
predictive power of perturbative calculations, even in applications
involving large momentum transfers, where $\alpha_{s}$ is
effectively small.

A number of theoretical approaches have been developed to
reorganize the perturbative expansions in an effort to improve the
predictability of the perturbative QCD (pQCD). For example,
optimized scale and scheme choices have been proposed, such as the
method of effective charges (ECH) [5], the principle of minimal
sensitivity (PMS) [6], and the Brodsky-Lepage-Mackenize (BLM)
scale-setting prescription [7] and its generalizations [8-20]. In
Ref.[4], the resummation of the formally divergent renormalon
series and the parametrization of related higher-twist
power-suppressed contributions has been given.

In general, a factorially divergent renormalon series arises when
one integrates over the logarithmically running coupling
$\alpha_{s}(k^2)$ in a loop diagram. Such contributions do not occur
in conformally invariant theories which have a constant coupling. Of
course, in the physical theory, the QCD coupling does run.

Among the fundamental predictions of QCD are asymptotic scaling
laws for large-angle exclusive processes [21-28]. QCD counting
rules were formalized in Refs.[22,23]. These reactions probe
hadronic constituents at large relative momenta, or equivalently,
the hadronic wave function at short distances. In particular, the
meson wave function determines the leading higher-twist
contribution to meson production at high $p_T$.

The hadronic wave functions in terms of quark and gluon degrees of
freedom play an important role in the quantum chromodynamics
predictions for hadronic processes. In the perturbative  QCD
theory, the hadronic distribution amplitudes and structure
functions which enter exclusive and inclusive processes via the
factorization theorems at high momentum transfers can be
determined by the hadronic wave functions, and therefore they are
the underlying links between hadronic phenomena in QCD at large
(nonperturbative) and small distances (perturbative). If the
hadronic wave functions were accurately known, then we could
calculate the hadronic distribution amplitude and structure
functions for exclusive and inclusive processes in QCD.

The study of hadron production at large transverse momenta in
hadronic interactions is a  valuable testing ground of the
perturbative regime of QCD, providing information on both the
parton distribution functions (PDFs) in the proton, and the
parton-to-hadron fragmentation functions (FFs)[29].

The frozen coupling constant approach can be applied for
investigation, not only for exclusive processes, but also for the
calculation of higher-twist contributions to some inclusive
processes, for example as large -$p_{T}$ meson photoproduction
[30], two-jet+meson production in the electron-positron
annihilation [31]. In the frozen coupling constant approach was
used Refs.[30,31] for calculation of integrals, such as
\begin{equation}
I\sim \int\frac{\alpha_{s}(\hat {Q}^2)\Phi(x,\hat{Q}^2)}{1-x}dx
\end{equation}
According to Ref.[7], in pQCD calculations, the argument
$\hat{Q}^2$ of the running coupling constant  should be taken
equal to the square of the momentum transfer of a hard gluon in a
corresponding Feynman diagram, in both the renormalization and
factorization scale. But defined in this way,
$\alpha_{s}(\hat{Q}^2)$ suffers from infrared singularities. For
example in Ref.[32], $\hat{Q}^2$ equals to $(x_{1}-1)\hat{u}$ and
$-x_{1}\hat{t}$, where $\hat{u}$, $\hat{t}$ are the subprocess's
Mandelstam invariants. Therefore, in the soft regions
$x_{1}\rightarrow 0$, $x_{2}\rightarrow 0$, the integrals in (1.1)
diverge and we need some regularization methods of
$\alpha_{s}(Q^2)$ in these regions for their calculation. In
Ref.[33], the authors investigated the phenomenology of infrared
renormalons in inclusive processes. The dispersive approach has
been devised to extend properly modified perturbation theory
calculations towards the low-energy region [34]. Connections
between power corrections for the three Deep Inelastic Scattering
sum rules have also been explored in Ref.[35].

Investigation of the infrared renormalon effects in various
inclusive and exclusive processes is one of the most important and
interesting problems in the perturbative QCD. It is known that
infrared renormalons are responsible for factorial growth of
coefficients in perturbative series for the physical quantities.
But, these divergent  series can be resummed by means of the Borel
transformation [1] and the principal  value prescription [36], and
effects of infrared renormalons can be taken into account by a
scale-setting procedure
$\alpha_{s}(Q^2)\rightarrow\alpha_{s}(exp(f(Q^2))Q^2)$ at the
one-loop order results. Technically, all-order resummation of
infrared renormalons corresponds to the calculation of the one-loop
Feynman diagrams with the running coupling constant
$\alpha_{s}(-k^2)$ at the vertices or, alternatively, to calculation
of the same diagrams with nonzero gluon mass. Studies of infrared
renormalon problems have also opened new prospects for evaluation of
power-suppressed corrections to processes characteristics [37].
Power corrections can also be obtained by means of the Landau-pole
free expression for the QCD coupling constant. The most simple and
elaborated variant of the dispersive approach, the Shirkov and
Solovtsov analytic perturbation theory, was formulated in Ref.[38].

By taking these points into account, it may be argued that the
analysis of the next-to-leading order higher-twist effects on the
dependence of the pion wave function in pion production at
proton-proton collisions by the running coupling approach, are
significant from both theoretical and experimental[39] points of
view.

In this work we apply the running coupling approach[40] in order
to compute the effects of the infrared renormalons on the meson
production in proton-proton collisions in next-to-leading order.
This approach was also employed previously[41,42] to calculate the
inclusive meson production in proton-proton and photon-photon
collisions. The running coupling method in next-to-leading order
pion and kaon electromagnetic form factor was computed in [43].

As we know, power-suppressed contributions to exclusive processes
in QCD, which are commonly referred to as higher-twist
corrections. The higher-twist approximation  describes the
multiple scattering of a parton as power corrections to the
leading-twist cross section.

We will show that higher-twist terms contribute substantially to the
inclusive meson cross section at moderate transverse momenta. In
addition, we shall demonstrate that higher-twist reactions
necessarily dominate in the kinematic limit where the transverse
momentum approaches the phase-spase boundary.

A precise measurement of the inclusive charged pion production
cross section at $\sqrt s=62.4\,\,GeV$ and $\sqrt s=200\,\,GeV$ is
important for the proton-proton collisions program at  the
Relativistic Heavy Ion Collider (RHIC) at the Brookhaven National
Laboratory. Another important aspect of this study is the choice
of the meson model wave functions. In this respect, the
contribution of the higher-twist Feynman diagrams to a pion
production cross section in proton-proton collisions is been
computed by using various pion wave functions. Also, higher-twist
contributions which are calculated by the running coupling
constant and frozen coupling constant approaches are been
estimated and compared to each other. Within this context, this
paper is organized as follows: In Sec.\ref{ht}, we provide
formulas for the calculation of the contribution of the high twist
diagrams. In Sec. \ref{ir} we present formulas and an analysis of
the next-to-leading order higher-twist effects on the dependence
of the pion wave function by the running coupling constant
approach. In Sec. \ref{lt}, we provide formulas for the
calculation of the contribution of the leading-twist diagrams. In
Sec. \ref{results}, we give the numerical results for the cross
section and discuss the dependence of the cross section on the
pion wave functions. We present our conclusions in Sec.
\ref{conc}.

\section{CONTRIBUTION OF THE HIGH TWIST DIAGRAMS}
\label{ht} The higher-twist Feynman diagrams, which describe the
subprocess $q_1+\bar{q}_{2} \to \pi^{+}(\pi^{-})+\gamma$ for the
pion production in the proton-proton collision are shown in Fig.1.
In the higher-twist diagrams, the pion of a proton quark is
directly observed. Their $1/Q^2$ power suppression is caused by a
hard gluon exchange between pion constituents. The amplitude for
this subprocess can be found by means of the Brodsky-Lepage
formula [27]:

\begin{equation}
M(\hat s,\hat
t)=\int_{0}^{1}{dx_1}\int_{0}^{1}dx_2\delta(1-x_1-x_2)\Phi_{\pi}(x_1,x_2,Q^2)T_{H}(\hat
s,\hat t;x_1,x_2).
\end{equation}

In Eq.(2.1), $T_H$ is  the sum of the graphs contributing to the
hard-scattering part of the subprocess. The hard-scattering part for
the subprocess under consideration is $q_1+\bar{q}_{2} \to
(q_{1}\bar{q}_2)+\gamma$, where a quark and antiquark form a
pseudoscalar, color-singlet state $(q_1\bar{q}_2)$. Here
$\Phi(x_1,x_2,Q^2)$ is the pion wave function, i.e., the probability
amplitude for finding the valence $q_1\bar{q}_2$ Fock state in the
meson carry fractions $x_1$ and $x_2$, $x_1+x_2=1$. Remarkably, this
factorization is gauge invariant and only requires that the momentum
transfers in $T_H$ be large compared to the intrinsic mass scales of
QCD. Since the distribution amplitude and the hard-scattering
amplitude are defined without reference to the perturbation theory,
the factorization is valid to leading order in $1/Q$, independent of
the convergence of perturbative expansions.

The hard-scattering amplitude $T_H$ can be calculated in
perturbation theory and represented as a series in the QCD running
coupling constant $\alpha_s(Q^2)$. The function $\Phi$ is
intrinsically nonperturbative, but its evolution can be calculated
perturbatively. In our calculation, we have neglected the pion and
the proton masses. Turning to extracting the contributions of the
higher-twist subprocesses, there are many kinds of leading-twist
subprocesses in $pp$ collisions as the background of the
higher-twist subprocess $q_1+q_2 \to \pi^{+}(or\,\,
\pi^-)+\gamma$, such as $q+\bar{q} \to \gamma+g(g \to
\pi^{+}(\pi^{-}))$, $q+g \to \gamma+q(q \to \pi^{+}(\pi^{-}))$,
$\bar{q}+g \to \gamma+\bar{q}g(\bar{q} \to \pi^{+}(\pi^{-}))$ etc.
The contributions from these leading-twist subprocesses strongly
depend on some phenomenological factors, for example, quark and
gluon distribution functions in the proton and fragmentation
functions of various constituents, etc. Most of these factors have
not been well determined, neither theoretically nor
experimentally. Thus they cause very large uncertainty in the
computation of the cross section of process $pp \to \pi^{+}(or\,\,
\pi^{-})+\gamma +X$. In general, the magnitude of this uncertainty
is much larger than the sum of all the higher-twist contributions,
so it is very difficult to extract the higher-twist contributions.

The production of a hadron at large transverse momentum, $p_T$, in
a hadronic collisions is conventionally analyzed within the
framework of perturbative QCD by convoluting the leading-twist $2
\to 2$ hard subprocess cross sections with evolved structure  and
fragmentation functions. The most important discriminant of the
twist of a perturbative QCD subprocess in a hard hadronic
collision is the scaling of the inclusive invariant cross section
[22-25,44],

\begin{equation}
\sigma^{inv}\equiv E\frac{d\sigma}{d^{3}p}(AB \to CX)=\frac{F(x_{p_{T}},\vartheta)}{p_{T}^{n}}
\end{equation}
at fixed $x_{T}=2p_{T}/\sqrt s$ and center-of-mass angle
$\vartheta$. As we know in the original parton model [45] the
power fall-off is simply $n=4$ since the $2 \to 2$ subprocess
amplitude for point-like partons is scale invariant, and there is
no dimensional parameter  as in a conformal theory. The detected
hadron $C$ can be produced directly in the hard subprocess
reaction as in an exclusive reaction, then such direct
higher-twist processes can give a significant contribution since
there is no suppression from jet fragmentation at large momentum
fraction carried by the hadron, $z$, and the trigger hadron is
produced without any waste of energy[44].

The Mandelstam invariant variables for subprocesses $q_1+\bar{q}_{2} \to \pi^{+}(\pi^{-})+\gamma$ are defined as
\begin{equation}
\hat s=(p_1+p_2)^2,\quad \hat t=(p_1-p_{\pi})^2,\quad \hat
u=(p_1-p_{\gamma})^2.
\end{equation}

In our calculation, we have also neglected the quark masses. We
have aimed to calculate the pion production cross section and to
fix the differences due to the use of various pion model
functions. We have used five different wave functions: the
asymptotic wave function (ASY), the Chernyak-Zhitnitsky wave
function [28,46], the CLEO wave function [47], the Braun-Filyanov
pion wave functions [48] and  the Bakulev-Mikhailov-Stefanis pion
wave function [49]. It should be noted that the wave functions of
pions also are developed in Refs.[50-52] by the Dubna group:

$$
\Phi_{CZ}(x,\mu_{0}^2)=\Phi_{asy}(x)\left[C_{0}^{3/2}(2x-1)+\frac{2}{3}C_{2}^{3/2}(2x-1)\right],
$$
$$
\Phi_{CLEO}(x,\mu_{0}^2)=\Phi_{asy}(x)\left[C_{0}^{3/2}(2x-1)+0.27C_{2}^{3/2}(2x-1)-0.22C_{4}^{3/2}(2x-1)\right],
$$
$$
\Phi_{BF}(x,\mu_{0}^2)=\Phi_{asy}(x)\left[C_{0}^{3/2}(2x-1)+0.44C_{2}^{3/2}(2x-1)+0.25C_{4}^{3/2}(2x-1)\right],
$$
\begin{equation}
\Phi_{BMS}(x,\mu_{0}^2)=\Phi_{asy}(x)\left[C_{0}^{3/2}(2x-1)+0.188C_{2}^{3/2}(2x-1)-0.13C_{4}^{3/2}(2x-1)\right],
\end{equation}
$$
\Phi_{asy}(x)=\sqrt{3}f_{\pi}x(1-x),
$$
$$
C_{0}^{3/2}(2x-1)=1,\,\,C_{2}^{3/2}(2x-1)=\frac{3}{2}(5(2x-1)^2-1),
$$
$$
C_{4}^{3/2}(2x-1)=\frac{15}{8}(21(2x-1)^4-14(2x-1)^2+1),
$$
where $f_{\pi}=0.923 GeV$ is the pion decay constant. Here, we have
denoted by $x\equiv x_1$, the longitudinal fractional momentum
carried by the quark within the meson. Then, $x_2=1-x$ and
$x_1-x_2=2x-1$. The pion wave function is symmetric under the
replacement $x_1-x_2\leftrightarrow x_2-x_1$.

Several important nonperturbative tools have been developed which
allow specific predictions for the hadronic wave functions
directly from theory and experiments. The QCD sum-rule technique
and lattice gauge theory provide constraints on the moments of the
hadronic distribution amplitude. However, the correct pion wave
function is still an open problem in QCD. It is known that the
pion wave function can be expanded over the eigenfunctions of the
one-loop Brodsky-Lepage equation, \emph{i.e.}, in terms of the
Gegenbauer polynomials $\{C_{n}^{3/2}(2x-1)\}:$

\begin{equation}
\Phi_{\pi}(x,Q^2)=\Phi_{asy}(x)\left[1+\sum_{n=2,4..}^{\infty}a_{n}(Q^2)C_{n}^{3/2}(2x-1)\right].
\end{equation}

The evolution of the wave function on the factorization scale $Q^2$
is governed by the functions $a_n(Q^2)$,
\begin {equation}
a_n(Q^2)=a_n(\mu_{0}^2)\left[\frac{\alpha_{s}(Q^2)}{\alpha_{s}(\mu_{0}^2)}\right]^{\gamma_n/\beta_0},
\end{equation}
$$
\frac{\gamma_2}{\beta_{0}}=\frac{50}{81},\,\,\,\frac{\gamma_4}{\beta_{0}}=\frac{364}{405},\,\,
n_f=3.
$$

 In Eq.(2.6), $\{\gamma_n\}$ are anomalous dimensions defined by
the expression,

\begin{equation}
\gamma_n=C_F\left[1-\frac{2}{(n+1)(n+2)}+4\sum_{j=2}^{n+1}
\frac{1}{j}\right].
\end{equation}
The constants $a_n(\mu_{0}^2)=a_{n}^0$ are input parameters that
form the shape of the wave functions and which can be extracted
from experimental data or obtained from the nonperturbative QCD
computations at the normalization point $\mu_{0}^2$. The QCD
coupling constant $\alpha_{s}(Q^2)$ at the one-loop approximation
is given by the expression,

\begin{equation}
\alpha_{s}(Q^2)=\frac{4\pi}{\beta_0 ln(Q^2/\Lambda^2)}.
\end{equation}
Here, $\Lambda$ is the fundamental QCD scale parameter, $\beta_0$ is
the QCD beta function one-loop coefficient, respectively,
$$
\beta_0=11-\frac{2}{3}n_f.
$$
The cross section for the higher-twist subprocess $q_1\bar{q}_{2}
\to \pi^{+}(\pi^{-})\gamma$ is given by the expression
\begin{equation}
\frac{d\sigma}{d\hat t}(\hat s,\hat t,\hat u)=\frac
{8\pi^2\alpha_{E} C_F}{27}\frac{\left[D(\hat t,\hat
u)\right]^2}{{\hat s}^3}\left[\frac{1}{{\hat u}^2}+\frac{1}{{\hat
t}^2}\right],
\end{equation}
where
\begin{equation}
D(\hat t,\hat u)=e_1\hat
t\int_{0}^{1}dx\left[\frac{\alpha_{s}(\hat Q_1^2)\Phi_{\pi}(x,\hat Q_1^2)}{1-x}\right]+e_2\hat
u\int_{0}^{1}dx\left[\frac{\alpha_{s}(\hat Q_2^2)\Phi_{\pi}(x,\hat Q_2^2)}{1-x}\right].
\end{equation}
Here $\hat Q_{1}^2=(x_1-1)\hat u,\,\,\,\,$and $\hat
Q_{2}^2=-x_1\hat t$,\,\, represent the momentum squared carried by
the hard gluon in Fig.1, $e_1(e_2)$ is the charge of
$q_1(\overline{q}_2)$ and $C_F=\frac{4}{3}$. The higher-twist
contribution to the large-$p_{T}$ pion production cross section in
the process $pp\to\pi^{+}(\pi^{-})+\gamma+X$ is [53]:
\begin{equation}
\Sigma_{M}^{HT}\equiv E\frac{d\sigma}{d^3p}=\int_{0}^{1}\int_{0}^{1}
dx_1 dx_2 G_{{q_{1}}/{h_{1}}}(x_{1})
G_{{q_{2}}/{h_{2}}}(x_{2})\frac{\hat s}{\pi} \frac{d\sigma}{d\hat
t}(q\overline{q}\to \pi\gamma)\delta(\hat s+\hat t+\hat u).
\end{equation}
$$\pi E\frac{d\sigma}{d^3p}=\frac{d\sigma}{dydp_{T}^2},$$
$$\hat s=x_1x_2s,$$
$$\hat t=x_1t,$$
\begin{equation}
\hat u=x_2u,
\end{equation}
$$t= -m_T \sqrt{s} e^{-y}=-p_T \sqrt{s}e^{-y},$$
$$u= -m_T \sqrt {s} e^y=-p_T \sqrt{s}e^{y},$$
$$
x_1=-\frac{x_{2}u}{x_{2}s+t}=\frac{x_{2}p_{T}\sqrt s
e^{y}}{x_{2}s-p_{T}\sqrt s e^{-y}},
$$
$$
x_2=-\frac{x_{1}t}{x_{1}s+u}=\frac{x_{1}p_{T}\sqrt s
e^{-y}}{x_{1}s-p_{T}\sqrt s e^{y}},
$$

where $m_T$ -- is the transverse mass of pion, which is given by
$$m_T^2=m^2+p_T^2$$

Let us first consider the frozen coupling approach. In this
approach, we take the four-momentum  square $\hat{Q}_{1,2}^2$ of
the hard gluon to be equal the pion's transverse momentum square
$\hat{Q}_{1,2}^2=p_{T}^2$. In this case, the QCD coupling constant
$\alpha_s$ in the integral (2.10) does not depend on the
integration variable. After this substitution calculation of
integral (2.10) becomes easy. Hence, the effective cross section
obtained after substitution of the integral (2.10) into the
expression (2.9) is referred as the frozen coupling effective
cross section. We will denote the higher-twist cross section
obtained using the frozen coupling constant approximation by
$(\Sigma_{\pi}^{HT})^0$.

We have extracted the following higher-twist subprocesses
contributing to the two covariant cross sections in Eq.(2.11)
\begin{equation}
\frac{{d\sigma}^1}{d\hat t}(u\bar{d} \to \pi^{+}\gamma), \,\,\,\,
\frac{{d\sigma}^2}{d\hat t}(\bar{d}u \to \pi^{+}\gamma), \,\,\,\,
\frac{{d\sigma}^3}{d\hat t}(\bar{u}d \to \pi^{-}\gamma), \,\,\,\,
\frac{{d\sigma}^4}{d\hat t}(d\bar{u} \to \pi^{-}\gamma).
\end{equation}
By charge conjugation invariance, we also have
\begin{equation}
\frac{{d\sigma}^1}{d\hat t}(u\bar{d} \to \pi^{+}\gamma)=\frac{{d\sigma}^3}{d\hat t}(\bar{u}d \to
\pi^{-}\gamma),\,\,
{\mbox {and}} \,\,\,
\frac{{d\sigma}^2}{d\hat t}(\bar{d}u\to \pi^{+}\gamma)= \frac{{d\sigma}^4}{d\hat t}(d\bar{u} \to
\pi^{-}\gamma).
\end{equation}

\section{THE RUNNING COUPLING APPROACH AND HIGHER-TWIST MECHANISM}\label{ir}
In this section, we shall calculate the integral (2.10) using the
running coupling constant method and also discuss the problem of
normalization of the higher-twist process cross section in the
context of the same approach.

As  is seen from (2.10), in general, one has to take into account
not only the dependence of $\alpha(\hat {Q}_{1,2}^2)$ on the scale
$\hat {Q}_{1,2}^2$, but also an evolution of $\Phi(x,\hat
{Q}_{1,2}^2)$ with $\hat {Q}_{1,2}^2$. The meson wave function
evolves in accordance with a Bethe-Salpeter-type equation.
Therefore, it is worth noting that, the renormalization scale
(argument of $\alpha_s$) should be equal to $Q_{1}^2=(x_1-1)\hat
u$, $Q_{2}^2=-x_1\hat t$, whereas the factorization scale [$Q^2$
in $\Phi_{M}(x,Q^2)$] is taken independent from $x$, we assume
$Q^2=p_{T}^2$. Such a approximation does not considerably change
the numerical results, but the phenomenon considered in this
article (effect of infrared renormalons) becomes transparent. The
main problem in our investigation is the calculation of the
integral in (2.10) by the running coupling constant approach. This
integral in the framework of the running coupling approach takes
the form

\begin{equation}
I(\mu_{R_{0}}^2)=\int_{0}^{1}\frac{\alpha_{s}(\lambda
\mu_{R_0}^2)\Phi_{M}(x,\mu_{F}^2)dx}{1-x}.
\end{equation}

The $\alpha_{s}(\lambda \mu_{R_0}^2)$ has the infrared singularity
at $x\rightarrow1$, if $\lambda=1-x$  or $x\rightarrow0$, if
$\lambda=x$ and  as a result integral $(3.1)$ diverges (the pole
associated with the denominator of the integrand is fictitious,
because $\Phi_{M}\sim(1-x)$, and therefore, the singularity of the
integrand at $x=1$  is caused only by $\alpha_{s}(\lambda
\mu_{R_0}^2)$). For the regularization of the integral, we express
the running coupling at scaling variable $\alpha_{s}(\lambda
\mu_{R_0}^2)$ with the aid of the renormalization group equation
in terms of the fixed one $\alpha_{s}(Q^2)$. The renormalization
group equation for the running coupling
$\alpha\equiv\alpha_{s}/\pi$ has the form [36]

\begin{equation}
\frac{\partial\alpha(\lambda Q^2)}{\partial
ln\lambda}\simeq-\frac{\beta_{0}}{4}[\alpha(\lambda
Q^2)]^2-\frac{\beta_{1}}{16}[\alpha(\lambda Q^2)]^3
\end{equation}
where
$$
\beta_{0}=11-\frac{2}{3}n_{f}\,\,\,\beta_{1}=102-\frac{38}{3}n_{f},
$$
The solution of Eq.(3.2), with the initial condition
$$
\alpha(\lambda)|_{\lambda=1}=\alpha\equiv\alpha_{s}(Q^2)/\pi,
$$
is [35]

\begin{equation}
\frac{\alpha(\lambda)}{\alpha}=\left[1+\alpha
\frac{\beta_{0}}{4}\ln{\lambda}-\frac{\alpha
\beta_{1}}{4\beta_{0}}\left(\frac{\ln
\alpha(\lambda)}{\alpha}-\ln{\frac{4\beta_{0}/\beta_{1}+\alpha(\lambda)}{4\beta_{0}/\beta_{1}+\alpha}}\right)\right]^{-1}.
\end{equation}

This transcendental equation can be solved iteratively by keeping
the leading $\alpha^kln^k\lambda$  and next-to-leading
$\alpha^kln^{k-1}\lambda$ powers. For $\lambda=(1-x)$ these terms
are given by
\begin{equation}
\alpha((1-x)Q^2)=\frac{\alpha_{s}}{1+\ln{\lambda/t}}-
\frac{\alpha_{s}^2\beta_{1}}{4\pi\beta_{0}}\frac{\ln[1+\ln\lambda/t]}{[1+\ln\lambda/t]^2}
\end{equation}

The first term in Eq.(3.4) is the solution of the renormalization
group Eq.(3.2) with leading power accuracy, whereas the whole
expression (3.4) is the solution of Eq.(3.2) with next-to-leading
power accuracy. After substituting Eq.(3.4) into Eq.(2.10) we get:
$$
D(\hat t,\hat u)=e_{1}\hat t\int_{0}^{1}dx\frac{\alpha_{s}(\lambda
\mu_{R_0}^2)\Phi_{M}(x,p_{T}^2)}{1-x}+ e_{2}\hat
u\int_{0}^{1}dx\frac{\alpha_{s}(\lambda
\mu_{R_0}^2)\Phi_{M}(x,p_{T}^2)}{1-x}=
$$
$$
e_{1}\hat t\alpha_{s}(-\hat u)\int_{0}^{1}dx_{1}
\frac{\Phi_{M}(x,p_{T}^2)}{(1-x)(1+\ln\lambda/t_{1})}- e_{1}\hat t
\frac{\alpha_{s}^{2}(-\hat u)\beta_{1}}{4\pi\beta_{0}}\int_{0}^{1}dx\frac{\Phi_{M}(x,p_{T}^2)\ln(1+\ln\lambda/t_{1})}{(1-x)(1+\ln\lambda/t_{1})^2}+
$$
$$
e_{2}\hat u\alpha_{s}(-\hat t)\int_{0}^{1}dx_{1}
\frac{\Phi_{M}(x,p_{T}^2)}{(1-x)(1+\ln\lambda/t_{2})}- e_{2}\hat u
\frac{\alpha_{s}^2(-\hat t)\beta_{1}}{4\pi\beta_{0}}\int_{0}^{1}dx\frac{\Phi_{M}(x,p_{T}^2)\ln(1+\ln\lambda/t_{2})}{(1-x)(1+\ln\lambda/t_{2})^2}=
$$
$$
e_{1}\hat t\alpha_{s}(-\hat u)\int_{0}^{1}dx
\frac{\Phi_{asy}(x)\left[1+\sum_{2,4,..}^{\infty}a_{n}(\mu_{0}^{2})
\left[\frac{\alpha_{s}(p_{T}^2)}{\alpha_{s}(\mu_{0}^2)}\right]^{\gamma_{n}/\beta_{0}}C_{n}^{3/2}(2x-1)\right]}{(1-x)
(1+\ln\lambda/t_{1})}-
$$
$$
e_{1}\hat t \frac{\alpha_{s}^{2}(-\hat u)\beta_{1}}{4\pi\beta_{0}}\int_{0}^{1}dx
\frac{\Phi_{asy}(x)\left[1+\sum_{2,4,..}^{\infty}a_{n}(\mu_{0}^{2})
\left[\frac{\alpha_{s}(p_{T}^2)}{\alpha_{s}(\mu_{0}^2)}\right]^{\gamma_{n}/\beta_{0}}C_{n}^{3/2}(2x-1)\right]
\ln(1+\ln\lambda/t_{1})}{(1-x)
(1+\ln\lambda/t_{1})^2}+
$$
$$
e_{2}\hat u\alpha_{s}(-\hat t)\int_{0}^{1}dx
\frac{\Phi_{asy}(x)\left[1+\sum_{2,4,..}^{\infty}a_{n}(\mu_{0}^{2})
\left[\frac{\alpha_{s}(Q^2)}{\alpha_{s}(\mu_{0}^2)}\right]^{\gamma_{n}/\beta_{0}}C_{n}^{3/2}(2x-1)\right]}{(1-x)
(1+\ln\lambda/t_{2})}-
$$
\begin{equation}
e_{2}\hat u\frac{\alpha_{s}^{2}(-\hat t)\beta_{1}}{4\pi\beta_{0}}\int_{0}^{1}dx
\frac{\Phi_{asy}(x)\left[1+\sum_{2,4,..}^{\infty}a_{n}(\mu_{0}^{2})
\left[\frac{\alpha_{s}(p_{T}^2)}{\alpha_{s}(\mu_{0}^2)}\right]^{\gamma_{n}/\beta_{0}}C_{n}^{3/2}(2x-1)\right]\ln(1+\ln\lambda/t_{2})}{(1-x)
(1+\ln\lambda/t_{2})^2}
\end{equation}

where $t_1=4\pi/\alpha_{s}(-\hat u)\beta_{0}$ and
$t_2=4\pi/\alpha_{s}(-\hat t)\beta_{0}$.

It should be note that in [41], we used $t_1 \simeq t_2$.

 The integral (3.5) is common and, of course, still divergent, but now it
is recast into a form, which is suitable for calculation. Using the
running coupling constant approach, this integral may be found as a
perturbative series in $\alpha_{s}(Q^2)$
\begin{equation}
D(\hat t,\hat u)\sim
\sum_{n=1}^{\infty}\left(\frac{\alpha_{s}}{4\pi}\right)^nS_{n}.
\end{equation}
The expression coefficients $S_n$ can be written as power series
in the number of light quark flavors or, equivalently, as a series
in power of $\beta_0$:
$$
S_{n}=C_{n}\beta_{0}^{n-1}
$$
The coefficients $C_n$ of this series demonstrate factorial growth
$C_n\sim(n-1)!$, which might indicate an infrared renormalon
nature of divergences in the integral (3.5) and corresponding
series (3.6). The procedure for dealing with such ill-defined
series is well known: one has to perform the Borel transform of
the series [15]
$$
B[D](u)=\sum_{n=0}^{\infty}\frac{D_n}{n!} u^n,
$$
then invert $B[D](u)$ to obtain the resummed expression (the Borel
sum) $D(\hat t,\hat u)$. After this we can find directly the resummed
expression for $D(Q^2)$. The change of the variable $x$ to
$z=\ln(1-x)$, as $\ln(1-x)=\ln\lambda$. Then
$$
D(\hat t,\hat u)=e_{1}\hat t \alpha_{s}(-\hat u) t_1\int_{0}^{1}dx
\frac{\Phi_{M}(x,p_{T}^2)}{(1-x)(t_1+z)}- e_{1}\hat t
\frac{\alpha_{s}^{2}(-\hat u) \beta_{1}t_{1}^2}
{4\pi\beta_{0}}\int_{0}^{1} dx \frac{\Phi_{M}(x,p_{T}^2)\ln(1+z/t_{1})}{(1-x)(t_1+z)^2}+
$$
\begin{equation}
e_{2}\hat u \alpha_{s}(-\hat t) t_2 \int_{0}^{1} dx
\frac{\Phi_{M}(x,p_{T}^2)}{(1-x)(t_2+z)}- e_{2}\hat u
\frac{\alpha_{s}^2(-\hat t)\beta_{1}t_{2}^2}{4\pi\beta_{0}}\int_{0}^{1}dx
\frac{\Phi_{M}(x,p_{T}^2)\ln(1+z/t_{2})}{(1-x)(t_2+z)^2}
\end{equation}

The basic theoretical problem is how to define the Borel sum $B[D](u)$ of the integral
in Eq.(3.7) for the quantities we are interested in. In QCD this problem is usually
solved using perturbative methods and calculating the corresponding multiloop Feynman
diagrams with a one-gluon line, dressed by the chains of fermion bubbles.

For the calculation the expression (3.7) we will apply the
integral representations of $1/(t+z)^\nu$ and $\ln(t+z)/(t+z)^2$
[54,55]. After this operation, formula (3.7) is simplified and we
can extract the Borel sum of the perturbative series (3.6) and the
corresponding Borel transform in dependence from the wave
functions of the meson, respectively. Also after such
manipulations the obtained expression can be used for numerical
computations.

 It is convenient to use the following integral representation for
$1/(t+z)^\nu$ and $\ln(t+z)/(t+z)^2:$
\begin{equation}
\frac{1}{(t+z)^\nu}=\frac{1}{\Gamma(\nu)}\int_{0}^{\infty}e^{-(t+z)u}u^{\nu-1}du,
Re\nu>0
\end{equation}
and
\begin{equation}
\frac{\ln(t+z)}{(t+z)^2}=\int_{0}^{\infty}e^{-(t+z)u}(1-C-\ln u)u du
\end{equation}
where $C\simeq 0.577216$ is the Euler-Mascheroni constant.

After inserting Eq.(3.8) and (3.9) into (3.7).
 then, we obtain
$$
D(\hat t,\hat u)=e_{1} \hat{t} \alpha_{s}(-\hat u) t_1 \int_{0}^{1}
\int_{0}^{\infty} \frac{\Phi_{M}(x,p_{T}^2)e^{-(t_1+z)u}du dx}{(1-x)}-
$$
$$
\frac{e_{1} \hat{t} \alpha_{s}^{2}(-\hat u)\beta_1 t_{1}^2}{4\pi\beta_0} \int_{0}^{1}
\int_{0}^{\infty} \frac{\Phi_{M}(x,p_{T}^2)e^{-(t_1+z)u}du dx}{(1-x)}+
$$
$$
e_{2} \hat{u} \alpha_{s}(-\hat t) t_2 \int_{0}^{1} \int_{0}^{\infty}
\frac{\Phi_{M}(x,p_{T}^2)e^{-(t_2+z)u}du dx}{(1-x)}-
$$
\begin{equation}
\frac{e_{2} \hat{u} \alpha_{s}^2(-\hat t) \beta_1t_{2}^2}{4\pi\beta_0} \int_{0}^{1} \int_{0}^{\infty}
\frac{\Phi_{M}(x,p_{T}^2)e^{-(t_2+z)u}du dx}{(1-x)}.
\end{equation}
In the case of $\Phi_{asy}(x)$ for $D(\hat t,\hat u)$, we get
$$
D(\hat t,\hat u)=\frac{4\sqrt{3}\pi f_{\pi} e_{1} \hat t}{\beta_{0}}\cdot \int_{0}^{\infty} du  e^{-t_{1}u}\left[\frac{1}{1-u}-
\frac{1}{2-u}\right]-\frac{4\sqrt{3} \pi f_{\pi} e_{1} \beta_{1} \hat
t}{\beta_{0}^3} \cdot
$$
$$
\int_{0}^{\infty}du e^{-t_{1}u}\left[\frac{1}{1-u}-
\frac{1}{2-u}\right](1-C-\ln u-\ln {t_1})u+
$$
$$
\frac{4\sqrt{3} \pi f_{\pi} e_{2} \hat
u}{\beta_{0}}\cdot \int_{0}^{\infty}due^{-t_{2}u}\left[\frac{1}{1-u}-
\frac{1}{2-u}\right]-\frac{4\sqrt{3} \pi f_{\pi} e_{2}\beta_{1} \hat
u}{\beta_{0}^3}\cdot
$$
\begin{equation}
\int_{0}^{\infty}due^{-t_{2}u}\left[\frac{1}{1-u}-
\frac{1}{2-u}\right](1-C-\ln u-\ln {t_2})u.
\end{equation}
In the case  of the $\Phi_{CZ}(x,Q^2)$ wave function, we find
$$
D(\hat t,\hat u)=\frac{4\sqrt{3}\pi f_{\pi}e_{1}\hat
t}{\beta_{0}}\cdot\int_{0}^{\infty}du e^{-t_{1}u}
\left[\frac{1}{1-u}-\frac{1}{2-u}+\right.
$$
$$
0.84\left[\frac{\alpha_{s}(p_{T}^2)}{\alpha_{s}(\mu_{0}^2)}\right]^{50/81}
\left[\frac{4}{1-u}-\frac{24}{2-u}+\frac{40}{3-u}-
\left.\frac{20}{4-u}\right]\right]-
$$
$$
\frac{4\sqrt{3}\pi f_{\pi}e_{1}{\beta_1}\hat
t}{\beta_{0}^3}\cdot \int_{0}^{\infty} du e^{-t_{1}u}
\left[\frac{1}{1-u}-\frac{1}{2-u} + \right.
$$
$$
0.84\left[\frac{\alpha_{s}(p_{T}^2)}{\alpha_{s}(\mu_{0}^2)}\right]^{50/81}
\left[\frac{4}{1-u}-\frac{24}{2-u}+\frac{40}{3-u}-
\left.\frac{20}{4-u}\right]\right](1-C-\ln u-\ln {t_1})u+
$$
$$
\frac{4\sqrt{3}\pi f_{\pi}e_{2}\hat
u}{\beta_{0}}\cdot\int_{0}^{\infty}du e^{-t_{2}u}
\left[\frac{1}{1-u}-\frac{1}{2-u}+\right.
$$
$$
0.84\left[\frac{\alpha_{s}(p_{T}^2)}{\alpha_{s}(\mu_{0}^2)}\right]^{50/81}
\left[\frac{4}{1-u}-\frac{24}{2-u}+\frac{40}{3-u}-
\left.\frac{20}{4-u}\right]\right]-
$$
$$
\frac{4\sqrt{3}\pi f_{\pi}e_{2}{\beta_1}\hat
u}{\beta_{0}^3}\int_{0}^{\infty}du e^{-t_{2}u}
\left[\frac{1}{1-u}-\frac{1}{2-u} + \right.
$$
\begin{equation}
0.84\left[\frac{\alpha_{s}(p_{T}^2)}{\alpha_{s}(\mu_{0}^2)}\right]^{50/81}
\left[\frac{4}{1-u}-\frac{24}{2-u}+\frac{40}{3-u}-
\left.\frac{20}{4-u}\right]\right](1-C-\ln u-\ln {t_2})u
\end{equation}

In the case of the $\Phi_{CLEO}(x,Q^2)$ wave function, we get
$$
D(\hat t,\hat u)=\frac{4\sqrt{3}\pi f_{\pi}e_{1}\hat
t}{\beta_{0}} \int_{0}^{\infty}du
e^{-t_{1}u}\left[\frac{1}{1-u}-\frac{1}{2-u}+ \right.
0.405\left[\frac{\alpha_{s}(p_{T}^2)}{\alpha_{s}(\mu_{0}^2)}\right]^{50/81}\cdot
$$
$$
\left[\frac{4}{1-u}-\frac{24}{2-u}+\frac{40}{3-u}-
\frac{20}{4-u}\right]-0.4125\left[\frac{\alpha_{s}(p_{T}^2)}{\alpha_{s}(\mu_{0}^2)}\right]^{364/405}\cdot
$$
$$
\left.
\left[\frac{8}{1-u}-\frac{120}{2-u}+\frac{560}{3-u}-\frac{1112}{4-u}+\frac{1008}{5-u}-\frac{336}{6-u}\right]\right]-
$$
$$
\frac{4\sqrt{3}\pi f_{\pi}e_{1}{\beta_1}\hat
t}{\beta_{0}^3} \int_{0}^{\infty}du
e^{-t_{1}u}\left[\frac{1}{1-u}-\frac{1}{2-u}+ \right.
0.405\left[\frac{\alpha_{s}(p_{T}^2)}{\alpha_{s}(\mu_{0}^2)}\right]^{50/81}\cdot
$$
$$
\left[\frac{4}{1-u}-\frac{24}{2-u}+\frac{40}{3-u}-
\frac{20}{4-u}\right]-0.4125\left[\frac{\alpha_{s}(Q^2)}{\alpha_{s}(\mu_{0}^2)}\right]^{364/405}\cdot
$$
$$
\left.
\left[\frac{8}{1-u}-\frac{120}{2-u}+\frac{560}{3-u}-\frac{1112}{4-u}+\frac{1008}{5-u}-\frac{336}{6-u}\right]\right]
(1-C-\ln u-\ln {t_1})u+
$$
$$
\frac{4\sqrt{3}\pi f_{\pi}e_{2}\hat
u}{\beta_{0}} \int_{0}^{\infty}du
e^{-t_{2}u}\left[\frac{1}{1-u}-\frac{1}{2-u}+ \right.
0.405\left[\frac{\alpha_{s}(p_{T}^2)}{\alpha_{s}(\mu_{0}^2)}\right]^{50/81}\cdot
$$
$$
\left[\frac{4}{1-u}-\frac{24}{2-u}+\frac{40}{3-u}-
\frac{20}{4-u}\right]-0.4125\left[\frac{\alpha_{s}(p_{T}^2)}{\alpha_{s}(\mu_{0}^2)}\right]^{364/405}\cdot
$$
$$
\left.
\left[\frac{8}{1-u}-\frac{120}{2-u}+\frac{560}{3-u}-\frac{1112}{4-u}+\frac{1008}{5-u}-\frac{336}{6-u}\right]\right]-
$$
$$
\frac{4\sqrt{3}\pi f_{\pi}e_{2}{\beta_1}\hat
u}{\beta_{0}^3} \int_{0}^{\infty}du
e^{-t_{2}u}\left[\frac{1}{1-u}-\frac{1}{2-u}+ \right.
0.405\left[\frac{\alpha_{s}(p_{T}^2)}{\alpha_{s}(\mu_{0}^2)}\right]^{50/81}\cdot
$$
$$
\left[\frac{4}{1-u}-\frac{24}{2-u}+\frac{40}{3-u}-
\frac{20}{4-u}\right]-0.4125\left[\frac{\alpha_{s}(p_{T}^2)}{\alpha_{s}(\mu_{0}^2)}\right]^{364/405}\cdot
$$
\begin{equation}
\left.
\left[\frac{8}{1-u}-\frac{120}{2-u}+\frac{560}{3-u}-\frac{1112}{4-u}+\frac{1008}{5-u}-\frac{336}{6-u}\right]\right]
(1-C-\ln u-\ln {t_2})u
\end{equation}

Also, in the case of the $\Phi_{BMS}(x,Q^2)$ wave function, we get
$$
D(\hat t,\hat u)=\frac{4\sqrt{3}\pi f_{\pi}e_{1}\hat
t}{\beta_{0}} \int_{0}^{\infty}du
e^{-t_{1}u}\left[\frac{1}{1-u}-\frac{1}{2-u}+ \right.
0.282\left[\frac{\alpha_{s}(p_{T}^2)}{\alpha_{s}(\mu_{0}^2)}\right]^{50/81}\cdot
$$
$$
\left[\frac{4}{1-u}-\frac{24}{2-u}+\frac{40}{3-u}-
\frac{20}{4-u}\right]-0.244\left[\frac{\alpha_{s}(p_{T}^2)}{\alpha_{s}(\mu_{0}^2)}\right]^{364/405}\cdot
$$
$$
\left.
\left[\frac{8}{1-u}-\frac{120}{2-u}+\frac{560}{3-u}-\frac{1112}{4-u}+\frac{1008}{5-u}-\frac{336}{6-u}\right]\right]-
$$
$$
\frac{4\sqrt{3}\pi f_{\pi}e_{1}{\beta_1}\hat
t}{\beta_{0}^3} \int_{0}^{\infty}du
e^{-t_{1}u}\left[\frac{1}{1-u}-\frac{1}{2-u}+ \right.
0.282\left[\frac{\alpha_{s}(p_{T}^2)}{\alpha_{s}(\mu_{0}^2)}\right]^{50/81}\cdot
$$
$$
\left[\frac{4}{1-u}-\frac{24}{2-u}+\frac{40}{3-u}-
\frac{20}{4-u}\right]-0.244\left[\frac{\alpha_{s}(p_{T}^2)}{\alpha_{s}(\mu_{0}^2)}\right]^{364/405}\cdot
$$
$$
\left.
\left[\frac{8}{1-u}-\frac{120}{2-u}+\frac{560}{3-u}-\frac{1112}{4-u}+\frac{1008}{5-u}-\frac{336}{6-u}\right]\right](1-C-\ln u-\ln {t_1})u+
$$
$$
\frac{4\sqrt{3}\pi f_{\pi}e_{2}\hat
u}{\beta_{0}} \int_{0}^{\infty}du
e^{-t_{2}u}\left[\frac{1}{1-u}-\frac{1}{2-u}+ \right.
0.282\left[\frac{\alpha_{s}(p_{T}^2)}{\alpha_{s}(\mu_{0}^2)}\right]^{50/81}\cdot
$$
$$
\left[\frac{4}{1-u}-\frac{24}{2-u}+\frac{40}{3-u}-
\frac{20}{4-u}\right]-0.244\left[\frac{\alpha_{s}(p_{T}^2)}{\alpha_{s}(\mu_{0}^2)}\right]^{364/405}\cdot
$$
$$
\left.
\left[\frac{8}{1-u}-\frac{120}{2-u}+\frac{560}{3-u}-\frac{1112}{4-u}+\frac{1008}{5-u}-\frac{336}{6-u}\right]\right]-
$$
$$
\frac{4\sqrt{3}\pi f_{\pi}e_{2}{\beta_1}\hat
u}{\beta_{0}^3} \int_{0}^{\infty}du
e^{-t_{2}u}\left[\frac{1}{1-u}-\frac{1}{2-u}+ \right.
0.282\left[\frac{\alpha_{s}(p_{T}^2)}{\alpha_{s}(\mu_{0}^2)}\right]^{50/81}\cdot
$$
$$
\left[\frac{4}{1-u}-\frac{24}{2-u}+\frac{40}{3-u}-
\frac{20}{4-u}\right]-0.244\left[\frac{\alpha_{s}(p_{T}^2)}{\alpha_{s}(\mu_{0}^2)}\right]^{364/405}\cdot
$$
\begin{equation}
\left.
\left[\frac{8}{1-u}-\frac{120}{2-u}+\frac{560}{3-u}-\frac{1112}{4-u}+\frac{1008}{5-u}-\frac{336}{6-u}\right]\right]
(1-C-\ln u-\ln {t_2})u
\end{equation}

Equations(3.11)-(3.14) is nothing more than the Borel sum of the
perturbative series (3.6), and the corresponding Borel transform in
the case $\Phi_{asy}(x)$ is
\begin{equation}
B[D](u)=\frac{1}{1-u}-\frac{1}{2-u},
\end{equation}
in the case $\Phi_{CZ}(x,Q^2)$ is
\begin{equation}
B[D](u)=\frac{1}{1-u}-\frac{1}{2-u}+0.84\left(\frac{\alpha_{s}(p_{T}^2)}{\alpha_{s}(\mu_{0}^2)}\right)^{50/81}
\left(\frac{4}{1-u}-\frac{24}{2-u}+\frac{40}{3-u}-\frac{20}{4-u}\right),
\end{equation}

 in the case $\Phi_{CLEO}(x,Q^2)$ is
$$
B[D](u)=\frac{1}{1-u}-\frac{1}{2-u}+0.405\left(\frac{\alpha_{s}(p_{T}^2)}{\alpha_{s}(\mu_{0}^2)}\right)^{50/81}
\left(\frac{4}{1-u}-\frac{24}{2-u}+\frac{40}{3-u}-\frac{20}{4-u}\right)-
$$
\begin{equation}
0.4125\left(\frac{\alpha_{s}(Q^2)}{\alpha_{s}(\mu_{0}^2)}\right)^{364/405}\left(\frac{8}{1-u}-
\frac{120}{2-u}+\frac{560}{3-u}-\frac{1112}{4-u}+\frac{1008}{5-u}-\frac{336}{6-u}\right).
\end{equation}
and in the case $\Phi_{BMS}(x,Q^2)$ is
$$
B[D](u)=\frac{1}{1-u}-\frac{1}{2-u}+0.282\left(\frac{\alpha_{s}(p_{T}^2)}{\alpha_{s}(\mu_{0}^2)}\right)^{50/81}
\left(\frac{4}{1-u}-\frac{24}{2-u}+\frac{40}{3-u}-\frac{20}{4-u}\right)-
$$
\begin{equation}
0.244\left(\frac{\alpha_{s}(p_{T}^2)}{\alpha_{s}(\mu_{0}^2)}\right)^{364/405}\left(\frac{8}{1-u}-
\frac{120}{2-u}+\frac{560}{3-u}-\frac{1112}{4-u}+\frac{1008}{5-u}-\frac{336}{6-u}\right).
\end{equation}
The series (3.6) can be recovered by means of the following formula
$$
C_{n}=\left(\frac{d}{du}\right)^{n-1}B[D](u)\mid_{u=0}
$$
The Borel transform $B[D](u)$ has poles on the real $u$ axis at
$u=1;2;3;4;5;6,$ which confirms our conclusion concerning the
infrared renormalon nature of divergences in (3.6). To remove them
from Eqs.(3.11)-(3.14) we applied the principal value prescription.

 Hence, the effective cross section obtained after substitution of the
expressions (3.10-3.14) into the expression (2.10) is referred as
the running coupling effective cross section. We will denote the
higher-twist cross section obtained using the running coupling
constant approach by $(\Sigma_{\pi}^{HT})^{res}.$

\section{CONTRIBUTION OF THE  LEADING-TWIST DIAGRAMS}\label{lt}
Regarding the higher-twist corrections to the pion production cross
section, a comparison of our results with leading-twist
contributions is crucial. We take two leading-twist subprocesses for
the pion production:(1) quark-antiquark annihilation $q\bar{q} \to
g\gamma$, in which the $\pi$ meson is indirectly emitted from the
gluon, $g \to \pi^{+}(\pi^{-})$ and (2) quark-gluon fusion, $qg \to
q\gamma $, with subsequent fragmentation of the final quark into a
meson, $q \to \pi^{+}(\pi^{-})$. The corresponding cross sections
are obtained in
\begin{equation}
\frac{d\sigma}{d\hat t}(q\bar{q} \to gq)=\frac{8}{9}\pi\alpha_E
\alpha_s(Q^2)\frac{e_{q}^2}{{\hat s}^2}\left(\frac{\hat t}{\hat
u}+\frac{\hat u}{\hat t}\right),
\end{equation}
\begin{equation}
\frac{d\sigma}{d\hat t}(qg \to q\gamma)=-\frac{\pi{e_{q}^2}\alpha_E
\alpha_s(Q^2)}{{3\hat s}^2}\left(\frac{\hat s}{\hat t}+\frac{\hat
t}{\hat s}\right).
\end{equation}
For the leading-twist contribution, we find
$$
\Sigma_{M}^{LT}\equiv E\frac{d\sigma}{d^3p}=\sum_{q}\int_{0}^{1}
dx_1 dx_2dz \left(G_{{q_{1}}/{h_{1}}}(x_{1})
G_{{q_{2}}/{h_{2}}}(x_{2})D_{g}^{\pi}(z)\frac{\hat s}{\pi
z^2}\frac{d\sigma}{d\hat t}(q\bar{q}\to g\gamma)+\right.
$$
\begin{equation}
\left.G_{{q_{1}}/{h_{1}}}(x_{1})
G_{{g}/{h_{2}}}(x_{2})D_{q}^{\pi}(z)\frac{\hat s}{\pi
z^2}\frac{d\sigma}{d\hat t}(qg \to q\gamma)\right) \delta(\hat
s+\hat t+\hat u),
\end{equation}
where
\begin{equation}
\hat s=x_{1}x_{2}s,\,\,\hat t=\frac{x_{1}t}{z},\,\,\hat
u=\frac{x_{2}u}{z},\,\, z=-\frac{x_{1}t+x_{2}u}{x_{1}x_{2}s}.
\end{equation}
$D_{g}^{\pi}(z)=D_{g}^{\pi^{+}}(z)=D_{g}^{\pi^{-}}(z)$ and
$D_{q}^{\pi}(z)$ represents gluon and quark  fragmentation functions
into a meson containing  gluon and quark of the same flavor. In the
leading-twist subprocess, the  $\pi$ meson is indirectly emitted
from the gluon and quark with the fractional momentum $z$. The
$\delta$ function can be expressed in terms of the parton kinematic
variables, and the $z$ integration can then be done. The final form
for the cross section is
$$
\Sigma_{M}^{LT}\equiv
E\frac{d\sigma}{d^3p}=\sum_{q}\int_{x_{1min}}^{1} dx_1
\int_{x_{2min}}^{1} dx_2 \left(G_{{q_{1}}/{h_{1}}}(x_{1})
G_{{q_{2}}/{h_{2}}}(x_{2})D_{g}^{\pi}(z) \cdot \frac{1}{\pi
z}\frac{d\sigma}{d\hat t}(q\bar{q} \to g\gamma)+\right.
$$
$$
\left.G_{{q_{1}}/{h_{1}}}(x_{1})
G_{{g}/{h_{2}}}(x_{2})D_{g}^{\pi}(z) \cdot \frac{1}{\pi
z}\frac{d\sigma}{d\hat t}(qg \to q\gamma)\right)=
$$
$$
\sum_{q}\int_{x_{1min}}^{1} dx_1 \int_{x_{2min}}^{1}
\frac{dx_2}{-(x_{1}t+x_{2}u)}\left(x_{1}G_{{q_{1}}/{h_{1}}}(x_{1})
sx_{2}G_{{q_{2}}/{h_{2}}}(x_{2})\frac{D_{g}^{\pi}(z)}{\pi}
\frac{d\sigma}{d\hat t}(q\overline{q} \to g\gamma)+\right.
$$
\begin{equation}
\left.x_{1}G_{{q_{1}}/{h_{1}}}(x_{1})
sx_{2}G_{{g}/{h_{2}}}(x_{2})\frac{D_{g}^{\pi}(z)}{\pi}\frac{d\sigma}{d\hat
t}(qg \to q\gamma)\right).
\end{equation}

\section{NUMERICAL RESULTS AND DISCUSSION}\label{results}

In this section, we discuss the numerical results for
next-to-leading order higher-twist effects with higher-twist
contributions calculated in the context of the running coupling
and frozen coupling approaches on the dependence of the chosen
meson wave functions in the process $pp \to \pi^{+}(or\,\,
\pi^{-})\gamma +X$. In the calculations, we use the asymptotic
wave function $\Phi_{asy}$, the Chernyak-Zhitnitsky $\Phi_{CZ}$,
the CLEO pion wave function [41], the Braun-Filyanov pion wave
functions [42], and the Bakulev-Mikhailov-Stefanis pion wave
function[ 43]. For the higher-twist subprocess, we take
$q_1+\bar{q}_{2} \to (q_1\bar{q}_2)+\gamma$ and we have extracted
the following four higher-twist subprocesses  contributing to
$pp\to \pi^{+}(or\,\,\pi^{-})\gamma$ cross sections: $u\bar{d} \to
\pi^{+}\gamma,$ $\bar{d}u \to \pi^{+}\gamma$, $\bar{u}d \to
\pi^{-}\gamma$, $d\bar{u} \to \pi^{-}\gamma$ contributing to cross
sections. For the dominant leading-twist subprocess for the pion
production, we take the quark-antiquark annihilation $q\bar{q} \to
g\gamma$, in which the $\pi$ meson is indirectly emitted from the
gluon and quark-gluon fusion, $qg \to q\gamma $, with subsequent
fragmentation of the final quark into a meson, $q \to
\pi^{+}(\pi^{-})$.  As an example for the quark distribution
function inside the proton, the MRST2003c package [56] has been
used. The higher twist subprocesses probe the meson wave functions
over a large range of $Q^2$ squared momentum transfer, carried by
the gluon. Therefore, in the diagram given in Fig.1 we take
$Q_{1}^2=(x_1-1){\hat u}$, $Q_{2}^2=-x_1\hat t$ , which we have
obtained directly from the higher-twist subprocesses diagrams. The
same $Q^2$ has been used as an argument of $\alpha_s(Q^2)$ in the
calculation of each diagram.

The results of our numerical calculations are plotted in
Figs.2-15. First of all, it is very interesting to compare the
resummed higher-twist cross sections with the ones obtained in the
framework of the frozen coupling approach. In Figs.2-4 we show the
dependence of next-to-leading order higher-twist cross sections
$(\Sigma_{\pi^{+}}^{HT})_{NLO}^{res}$ calculated in the context of
the running coupling constant approach and the ratios
$R=(\Sigma_{\pi^{+}}^{HT})_{NLO}^{res}$/$\Sigma_{\pi^{+}}^{HT})_{LO}^{0}$,
$(\Sigma_{\pi^{+}}^{HT})_{NLO}^{res}$/$(\Sigma_{\pi^{+}}^{HT})_{LO}^{res}$
as a function of the pion transverse momentum $p_{T}$ for
different pion wave functions at $y=0$. It is seen that the values
of cross sections $(\Sigma_{\pi^{+}}^{HT})_{NLO}^{res}$, and
ratios for fixed $y$ and $\sqrt s$ depend on the choice of the
pion wave function. As seen from Fig.2 the next-to-leading order
higher-twist differential cross section is monotically decreasing
with an increase in the transverse momentum of the pion. In
Figs.3-5, we show the dependence of the ratios
$R=(\Sigma_{\pi^{+}}^{HT})_{NLO}^{res}$/$\Sigma_{\pi^{+}}^{HT})_{LO}^{0}$,
$(\Sigma_{\pi^{+}}^{HT})_{NLO}^{res}$/$(\Sigma_{\pi^{+}}^{HT})_{LO}^{res}$,
$(\Sigma_{\pi^{+}}^{HT})_{NLO}^{res}$/$(\Sigma_{\pi^{+}}^{LT})$ as
a function of the pion transverse momentum $p_{T}$ for different
pion wave functions. Here $(\Sigma_{\pi^{+}}^{HT})_{NLO}^{res}$,
$(\Sigma_{\pi^{+}}^{HT})_{LO}^0$ ,$(\Sigma_{\pi^{+}}^{LT})$ are
the higher-twist cross sections calculated in the context of the
running coupling method, in the framework of the frozen coupling
approach and is the leading-twist cross section, respectively. As
seen from Fig.5, in the region $2\,\,GeV/c<p_T<4\,\,GeV/c$
next-to-leading order higher-twist cross section calculated in the
context of the running coupling method is suppressed by about 1-2
orders of magnitude relative to the leading-twist cross section,
but in the region $5\,\,GeV/c<p_T\leq 30\,\,GeV/c$ is comparable
with the cross section of leading-twist. In Figs.6-8, we have
depicted next-to-leading order higher-twist cross
sections$(\Sigma_{\pi^{+}}^{HT})_{NLO}^{res}$, ratios
$R=(\Sigma_{\pi^{+}}^{HT})_{NLO}^{res}/(\Sigma_{\pi^{+}}^{HT})_{LO}^{res}$,\,
$(\Sigma_{\pi^{+}}^{HT})_{NLO}^{res}/(\Sigma_{\pi^{+}}^{LT})$ as a
function of the rapidity $y$ of the pion at $\sqrt s=62.4\,\,GeV$
and $p_T=4.9\,\,GeV/c$. At $\sqrt s=62.4\,\,GeV$ and
$p_T=4.9\,\,GeV/c$, the pion rapidity lies in the region
$-2.52\leq y\leq2.52$.

As seen from Figs.6-8, next-to-leading order higher-twist cross
section and ratios have a different distinctive. In the region
($0.2\leq y\leq 2.52$) the next-to-leading order higher-twist
cross section  $(\Sigma_{\pi^{+}}^{HT})_{NLO}^{res}$, is
suppressed by about 1-2 order of magnitude relative to the
leading-twist cross section. In the region ($-2.52\leq y\leq
-1.92$), the ratio for all wave functions increase with an
increase of the $y$ rapidity of the pion and has a maximum
approximately at the point $y=-1.92$. Besides that, the ratio
decreases with an increase in the $y$ rapidity of the pion. As is
seen from Figs.7-8, the ratio $R$  very sensitive to the choice of
the meson wave functions. But, as seen from Fig.8, the ratio
$(\Sigma_{\pi^{+}}^{HT})_{NLO}^{res}$/$(\Sigma_{\pi^{+}}^{LT})$
for all wave functions has a minimum approximately at the point
$y=-1.92$. Also, the distinction between $R(\Phi_{asy}(x))$ with
$R(\Phi_{CLEO}(x,Q^2))$, $R(\Phi_{CZ}(x,Q^2))$,
$R(\Phi_{BF}(x,Q^2))$   and $R(\Phi_{BMS}(x,Q^2))$ have been
calculated. For example, in the case of $\sqrt s=62.4\,\,GeV$,
$y=0$, the distinction between $R(\Phi_{asy}(x))$ with
$R(\Phi_{i}(x,Q^2))$\,\,(i=CLEO, CZ, BF, BMS) as a function of the
pion transverse momentum $p_{T}$ is shown in Table \ref{table1}.
Thus, the distinction between $R(\Phi_{asy}(x))$ and
$R(\Phi_{i}(x,Q^2))(i=CLEO,CZ,BF)$ is maximum at
$p_T=20\,\,GeV/c$, with $R(\Phi_{BMS}(x))$  at  $p_T=2\,\,GeV/c$
but the distinction between $R(\Phi_{asy}(x))$ with
$R(\Phi_{i}(x,Q^2))(i=CLEO, CZ,BF)$ is minimum at
$p_T=2\,\,GeV/c$, with $R(\Phi_{BMS}(x))$  at $p_T=20\,\,GeV/c$
and increase with an increase in $p_T$. Such a behavior  of $R$
may be explained by reducing all moments of the pion model wave
functions to those of $\Phi_{asy}(x)$ for high $Q^2$. Also, we
have calculated the distinction between $R(\Phi_{asy}(x))$ with
$R(\Phi_{CLEO}(x,Q^2))$, $R(\Phi_{CZ}(x,Q^2))$,
$R(\Phi_{BF}(x,Q^2))$ and $R(\Phi_{BMS}(x,Q^2))$ as a function of
the rapidity $y$ of the pion. For example, in the case of $\sqrt
s=62.4GeV$, $p_{T}=4.9GeV/c$  the distinction between
$R(\Phi_{asy}(x))$ with $R(\Phi_{i}(x,Q^2))$\,\,(i=CLEO, CZ, BF,
BMS) as a function of the rapidity $y$ of the pion is presented in
Table \ref{table2}.

We have also carried out comparative calculations in the
center-of-mass energy $\sqrt s=200\,\,GeV$. The results of our
numerical calculations in the center-of-mass energies $\sqrt
s=200\,\,GeV$ are plotted in Figs.9-15. Analysis of our
calculations at the center-of-mass energies $\sqrt s=62.4\,\,GeV$
and $\sqrt s=200\,\,GeV$, show that with the increase in beam energy
values of the cross sections, ratio
$R=(\Sigma_{\pi^{+}}^{HT})_{NLO}^{res}/(\Sigma_{\pi^{+}}^{HT})_{LO}^{0}$, and
contributions of next-to-leading order higher-twist to the cross
section decrease by about 1-2 order. Therefore the experimental
investigation  of higher-twist effects include renormalon effects
conveniently in low energy.
 On the other hand, the higher-twist corrections and ratios $R$ a
very sensitive to the choice of the pion wave function.
Analysis of our calculations at the center-of-mass energies $\sqrt s=62.4\,\,GeV$
and $\sqrt s=200\,\,GeV$, show that  values of the next-to-leading
order cross sections decrease by about $(25-30)$ percent of magnitude
relative to the leading-order cross sections.
Also, the distinction between $R(\Phi_{asy}(x))$ with
$R(\Phi_{CLEO}(x,Q^2))$, $R(\Phi_{CZ}(x,Q^2))$,
$R(\Phi_{BF}(x,Q^2))$  and $R(\Phi_{BMS}(x,Q^2))$ have been
calculated. For example, in the case of $\sqrt s=200\,\,GeV$, $y=0$,
the distinction between $R(\Phi_{asy}(x))$ with
$R(\Phi_{i}(x,Q^2))$\,\,(i=CLEO, CZ, BF, BMS) as a function of the
pion transverse momentum $p_{T}$ is shown in Table \ref{table3}.
Thus, the distinction between $R(\Phi_{asy}(x))$ with
$R(\Phi_{i}(x,Q^2))$,\,\,(i= CLEO, BF,) is maximum at
$p_T=35\,\,GeV/c$, with $R(\Phi_{BMS}(x))$  at  $p_T=10\,\,GeV/c$,
but the distinction between $R(\Phi_{asy}(x))$ with
$R(\Phi_{CZ}(x,Q^2))$, $R(\Phi_{CLEO}(x,Q^2))$,
$R(\Phi_{BF}(x,Q^2))$ is minimum at $p_T=10\,\,GeV/c$, with
$R(\Phi_{BMS}(x))$ at $p_T=95\,\,GeV/c$  and increase with an
increase in $p_T$. Also, we have calculated the distinction between
$R(\Phi_{asy}(x))$ with $R(\Phi_{CLEO}(x,Q^2))$,
$R(\Phi_{CZ}(x,Q^2))$, $R(\Phi_{BF}(x,Q^2))$ and
$R(\Phi_{BMS}(x,Q^2))$as a function of the rapidity $y$ of the pion.
For example, in the case of $\sqrt s=200GeV$, $p_{T}=15.5GeV/c$  the
distinction between $R(\Phi_{asy}(x))$ with $R(\Phi_{i}(x,Q^2))$ as
a function of the rapidity $y$ of the pion is presented in Table
\ref{table4}. The calculations show that the ratio
$R(\Phi_{i}(x,Q^2))$/$R(\Phi_{asy}(x))$, (i=CLEO, CZ, BF, BMS) for
all values of the transverse momentum $p_T$ of the pion identically
equivalent to ratio $r(\Phi_{i}(x,Q^2))$/$r(\Phi_{asy}(x))$.

In our calculations of the next-to-leading order higher-twist cross
section of the process the dependence of the transverse momentum of
meson appears in the range of $(10^{-5}\div10^{-25})mb/GeV^2$.
Therefore, higher-twist cross section obtained in our paper should be
observable at RHIC

\section{Concluding Remarks}\label{conc}
In this work we have calculated the inclusive meson production via
higher-twist mechanism and obtained the expressions for the
subprocess $q\overline{q}\rightarrow M \gamma$ cross section for
mesons with symmetric wave functions. For calculation of the
next-to-leading order cross section we have applied the running
coupling constant method and revealed infrared renormalon poles in
the cross section expression. Infrared renormalon induced
divergences have been regularized by the means of the principal
value prescripton and the resummed expression (the Borel sum) for
the next-to-leading order higher-twist cross section has been
found. In the resummed higher-twist cross section differs
considerably from that found using the frozen coupling approach,
some regions. Also we have demonstrated that next-to-leading order
higher-twist contributions to meson production cross section in
the proton-proton collisions have important phenomenological
consequences. Our investigation enables us to conclude that the
higher-twist pion production cross section in the proton-proton
collisions depends on the form of the pion model wave functions
and may be used for their study. Analysis of our calculations
shows that the magnitude of next-to-leading order cross sections
calculated in the running coupling approach in some regions is
larger  than the leading-twist cross sections in 1-2 order.

Further investigations are needed in order to clarify the role of
high twist effects  in this process. We have demonstrated that the
resummed result depends on the pion model wave functions used in
calculations. The proton-proton collisions provide us with a new
opportunity to probe a proton's internal structure. In particular,
meson production in proton-proton collisions  takes into account
infrared renormalon effects: this opens a window  toward new types
of parton distributions which can not be measured
by the deep inelastic lepton-proton scatterings.
Finally, we discuss the phenomenological consequences of possible
higher-twist contributions to meson production in proton-proton
collisions in next-to-leading order at RHIC.
Future RHIC measurements will provide further tests of the dynamics of
large $p_T$ hadron production beyond leading twist.

\section*{Acknowledgments}
The work presented in this paper was completed while one of the authors,
A.I.Ahmadov, was visiting the HECAP section of the Abdus Salam ICTP, Trieste, Italy. He
would like to express his gratitude  to the members of the section especially
to the head Prof. S. Randjbar-Daemi  for their hospitality. Financial support
by ICTP is also gratefully acknowledged.

\newpage

\begin{table}[ht]
\begin{center}
\begin{tabular}{|c|c|c|c|c|c} \hline
$p_{T},GeV/c$ & $\frac{R(\Phi_{CLEO}(x,Q^2))}{R(\Phi_{asy}(x))}$ &
$\frac{R(\Phi_{CZ}(x,Q^2))}{R(\Phi_{asy}(x))}$ &
$\frac{R(\Phi_{BF}(x,Q^2))}{R(\Phi_{asy}(x))}$ &
$\frac{R(\Phi_{BMS}(x,Q^2))}{R(\Phi_{asy}(x))}$ \\
\hline
  2 & 0.391  & 0.212 & 0.338 & 7.628 \\ \hline
  6 & 1.236 & 0.387 & 1.03 &1.632 \\ \hline
  20 & 5.169  & 4.226 & 4.326& 2.268\\ \hline

\end{tabular}
\end{center}
\caption{The distinction between $R(\Phi_{asy}(x))$ with
$R(\Phi_{i}(x,Q^{2}))$  (i=CLEO, CZ, BF, BMS) at c.m.
 energy $\sqrt s=62.4\,\,GeV$.} \label{table1}
\end{table}

\begin{table}[ht]
\begin{center}
\begin{tabular}{|c|c|c|c|c|c} \hline
$y$ & $\frac{R(\Phi_{CLEO}(x,Q^2))}{R(\Phi_{asy}(x))}$ &
$\frac{R(\Phi_{CZ}(x,Q^2))}{R(\Phi_{asy}(x))}$ &
$\frac{R(\Phi_{BF}(x,Q^2))}{R(\Phi_{asy}(x))}$&
$\frac{R(\Phi_{BMS}(x,Q^2))}{R(\Phi_{asy}(x))}$ \\
\hline -2.52& 7.679 & 1.398 & 7.142 & 2.124 \\
\hline -1.92 &0.241 &0.226 & 0.313&3.486 \\
\hline 0.78& 0.051  & 0.637 & 4.367 & 1.849\\
\hline
\end{tabular}
\end{center}
\caption{The distinction between $R(\Phi_{asy}(x))$ with
$R(\Phi_{i}(x,Q^{2}))$ (i=CLEO, CZ, BF, BMS) at c.m. energy $\sqrt
s=62.4\,\,GeV$ and $p_T=4.9\,\, GeV/c$.} \label{table2}
\end{table}

\begin{table}[ht]
\begin{center}
\begin{tabular}{|c|c|c|c|c|c}                 \hline
$p_{T},GeV/c$ & $\frac{R(\Phi_{CLEO}(x,Q^2))}{R(\Phi_{asy}(x))}$ &
$\frac{R(\Phi_{CZ}(x,Q^2))}{R(\Phi_{asy}(x))}$ &
$\frac{R(\Phi_{BF}(x,Q^2))}{R(\Phi_{asy}(x))}$&
$\frac{R(\Phi_{BMS}(x,Q^2))}{R(\Phi_{asy}(x))}$ \\
\hline
  10 & 0.706  & 0.246 & 0.438 & 1.825\\ \hline
  35 & 2.746  & 0.824 & 2.746 & 0.684 \\ \hline
  95& 2.021  & 0.269 & 0,807 &0.342 \\ \hline
\end{tabular}
\end{center}
\caption{The distinction between $R(\Phi_{asy}(x))$ with
$R(\Phi_{i}(x,Q^{2}))$  (i=CLEO, CZ, BF, BMS) at c.m. energy $\sqrt
s=200\,\, GeV$.}\label{table3}
\end{table}

\newpage

\begin{table}[ht]
\begin{center}
\begin{tabular}{|c|c|c|c|c|c}\hline
$y$ & $\frac{r(\Phi_{CLEO}(x,Q^2))}{r(\Phi_{asy}(x))}$ &
$\frac{r(\Phi_{CZ}(x,Q^2))}{r(\Phi_{asy}(x))}$ &
$\frac{r(\Phi_{BF}(x,Q^2))}{r(\Phi_{asy}(x))}$ &
$\frac{R(\Phi_{BMS}(x,Q^2))}{R(\Phi_{asy}(x))}$ \\
\hline
  -2.52 & 3.468  & 0.564 & 2.121& 4.214 \\ \hline
  -1.92 & 1.068  & 0.204 & 0.267 &0.768 \\ \hline
  0.78& 0.328  & 0.579 & 3.322 &3.021\\ \hline
\end{tabular}
\end{center}
\caption{The distinction between $R(\Phi_{asy}(x))$ with
$R(\Phi_{i}(x,Q^{2}))$ (i=CLEO, CZ, BF, BMS) at c.m. energy $\sqrt
s=200\,\, GeV$ and $p_T=15.5\,\, GeV/c$.} \label{table4}
\end{table}

\newpage

\begin{figure}[htb]
\vskip 1.2cm \epsfxsize 16cm \centerline{\epsfbox{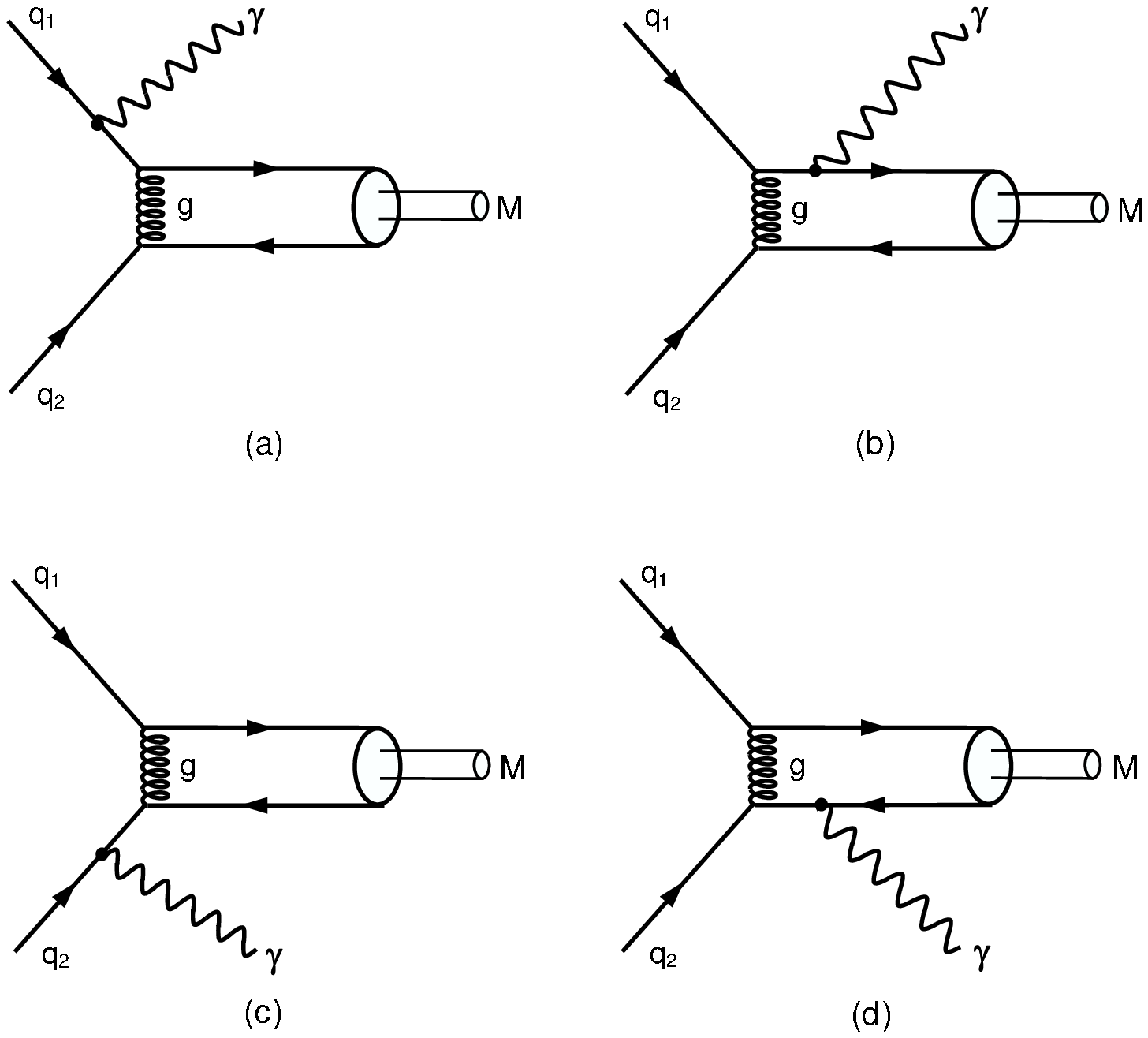}}
\vskip-5cm \caption{Feynman diagrams for the  higher-twist
subprocess, $q_1 q_2 \to \pi^{+}(or\,\,\pi^{-})\gamma.$}
\label{Fig1}
\end{figure}

\newpage

\begin{figure}[htb]
\vskip-1.2cm\epsfxsize 11.8cm \centerline{\epsfbox{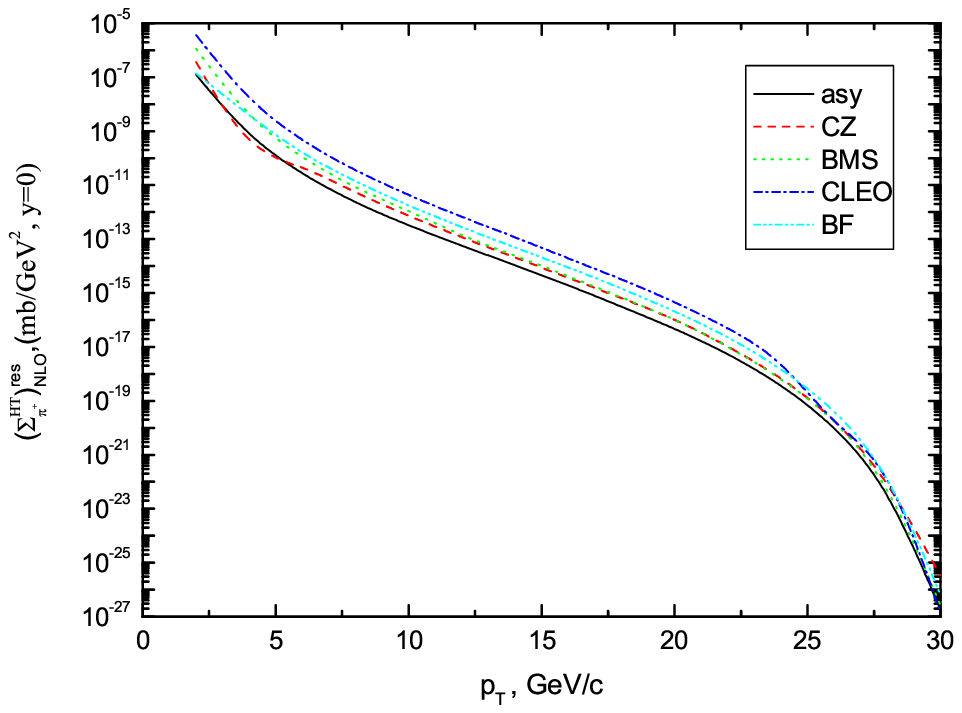}}
\vskip-0.2cm \caption{Next-to-leading order higher-twist $\pi^{+}$ production cross
section $(\Sigma_{\pi^{+}}^{HT})_{NLO}^{res}$ as a function of the $p_{T}$ transverse
momentum of the pion at the c.m.energy  $\sqrt s=62.4\,\,
GeV$.}\label{Fig2}
 \vskip-1.0cm
\vskip 1.8cm \epsfxsize 11.8cm \centerline{\epsfbox{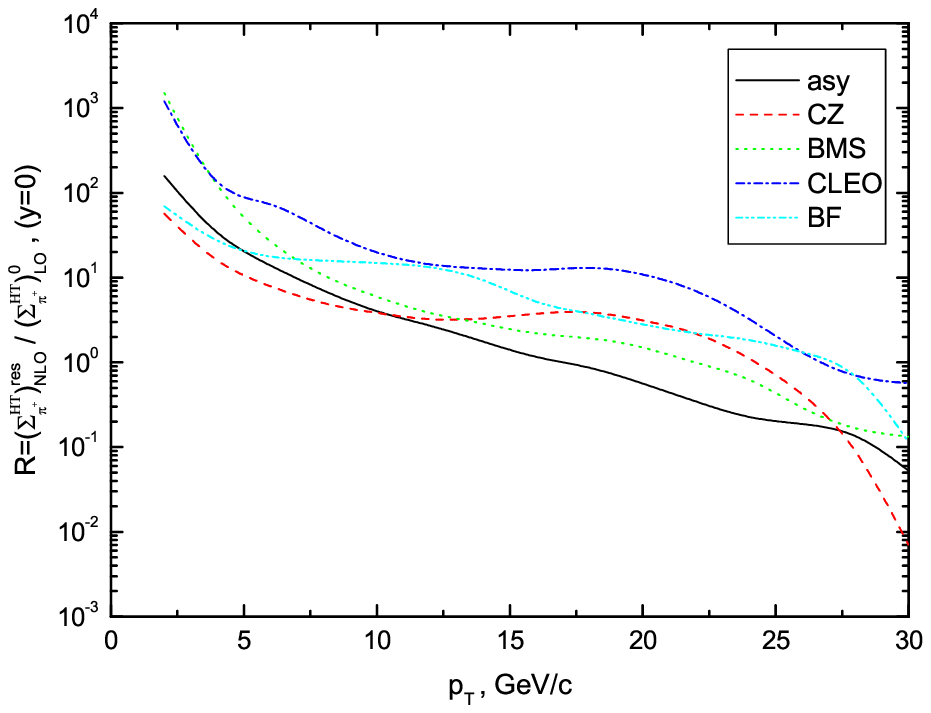}}
\vskip-0.05cm \caption{Ratio
$R=(\Sigma_{\pi^{+}}^{HT})_{NLO}^{res}/(\Sigma_{\pi^{+}}^{HT})_{LO}^{0}$, where
LO and NLO higher-twist contribution are calculated for the pion rapidity $y=0$
at the c.m.energy $\sqrt s=62.4\,\,GeV$ as a function of the pion
transverse momentum, $p_{T}$.} \label{Fig3}
\end{figure}

\newpage

\begin{figure}[htb]
 \vskip-1.2cm\epsfxsize 11.8cm \centerline{\epsfbox{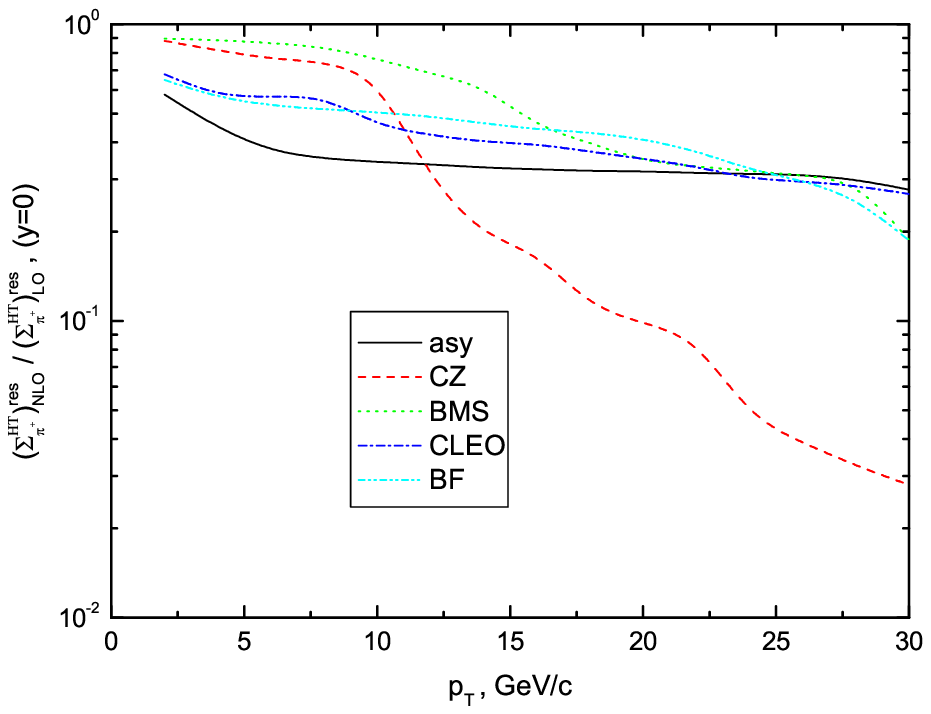}} \vskip-0.2cm
\caption{Ratio
$(\Sigma_{\pi^{+}}^{HT})_{NLO}^{res}/(\Sigma_{\pi^{+}}^{HT})_{LO}^{res}$, where
higher-twist contribution are calculated for the pion rapidity $y=0$
at the c.m.energy $\sqrt s=62.4\,\,GeV$ as a function of the pion
transverse momentum, $p_{T}$.}
\label{Fig4}
 \vskip 1.8cm
 \vskip-1.2cm\epsfxsize 11.8cm \centerline{\epsfbox{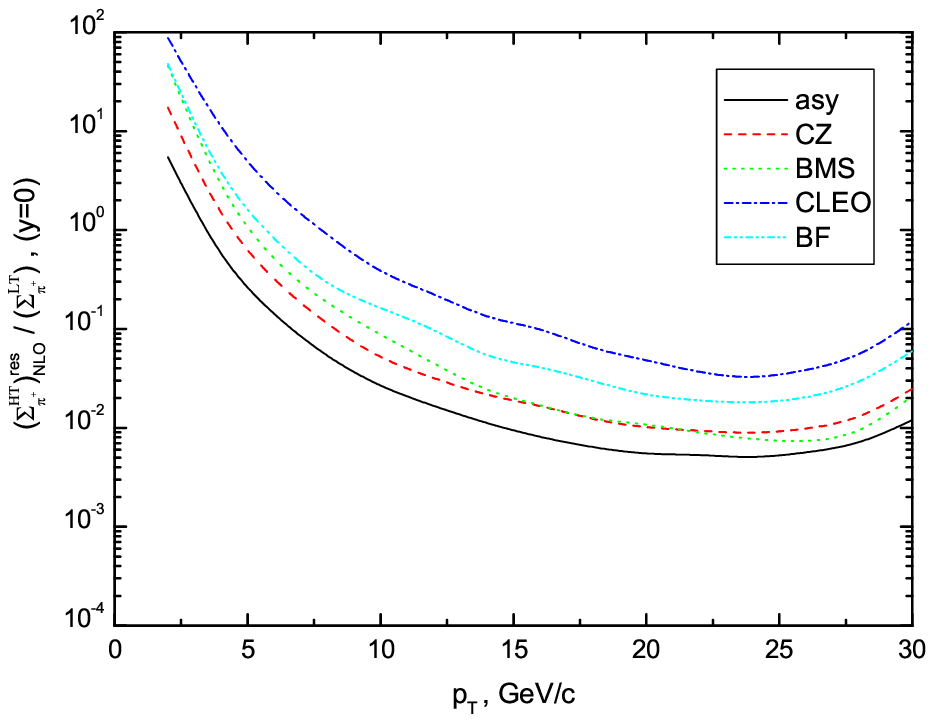}} \vskip-0.2cm
\caption{Ratio
$(\Sigma_{\pi^{+}}^{HT})_{NLO}^{res}/(\Sigma_{\pi^{+}}^{LT})$, as a function of the
$p_{T}$ transverse momentum of the pion
at the c.m.energy $\sqrt s=62.4\,\,GeV$.} \label{Fig5}
\end{figure}

\newpage

\begin{figure}[htb]
\vskip-1.2cm \epsfxsize 11.8cm \centerline{\epsfbox{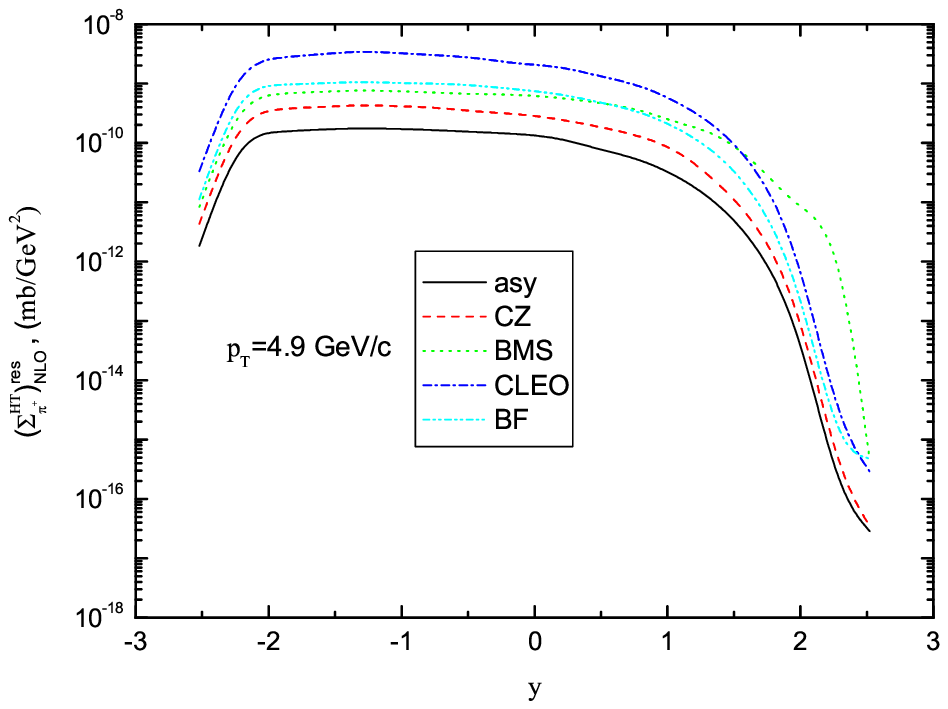}}
\vskip-0.2cm \caption{Next-to-leading order higher-twist $\pi^{+}$ production cross
section $(\Sigma_{\pi^{+}}^{HT})_{NLO}^{res}$, as a function of the $y$
rapidity of the pion at the  transverse momentum of the pion
$p_T=4.9\,\, GeV/c$, at the c.m. energy $\sqrt s=62.4\,\, GeV$.} \label{Fig6}
\vskip-1.0cm
\vskip 1.8cm\epsfxsize 11.8cm \centerline{\epsfbox{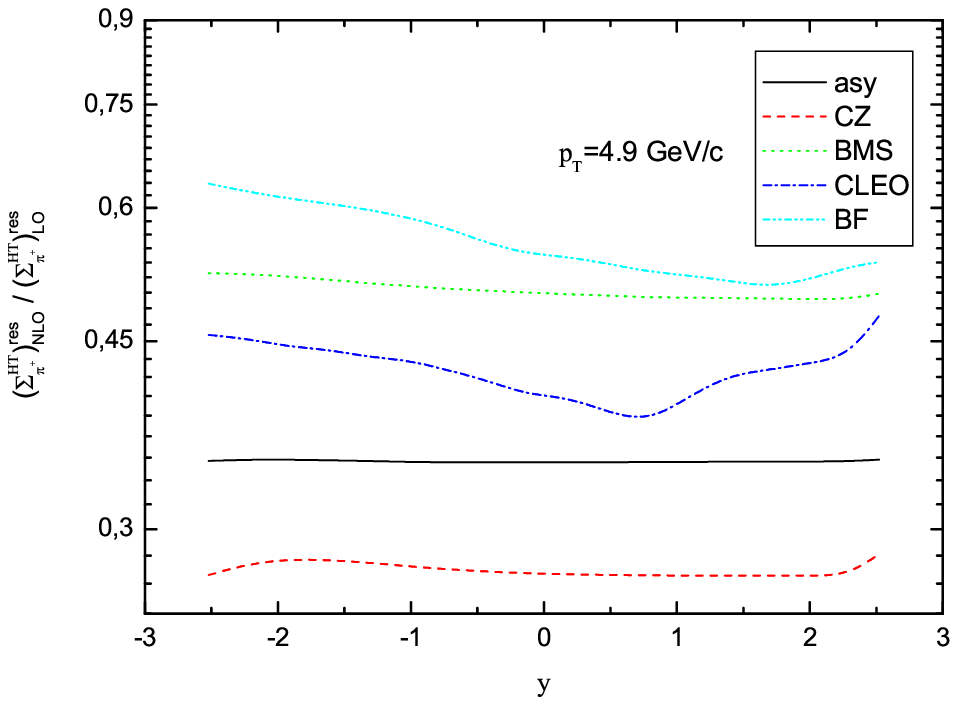}}
\vskip-0.05cm \caption{Ratio
$(\Sigma_{\pi^{+}}^{HT})_{NLO}^{res}/(\Sigma_{\pi^{+}}^{HT})_{LO}^{res}$,
as a function of the $y$ rapidity of the pion at the  transverse momentum
of the pion $p_T=4.9\,\, GeV/c$, at the c.m. energy $\sqrt
s=62.4\,\, GeV$.}
\label{Fig7}
\end{figure}

\newpage

\begin{figure}[htb]
\vskip-1.2cm \epsfxsize 11.8cm \centerline{\epsfbox{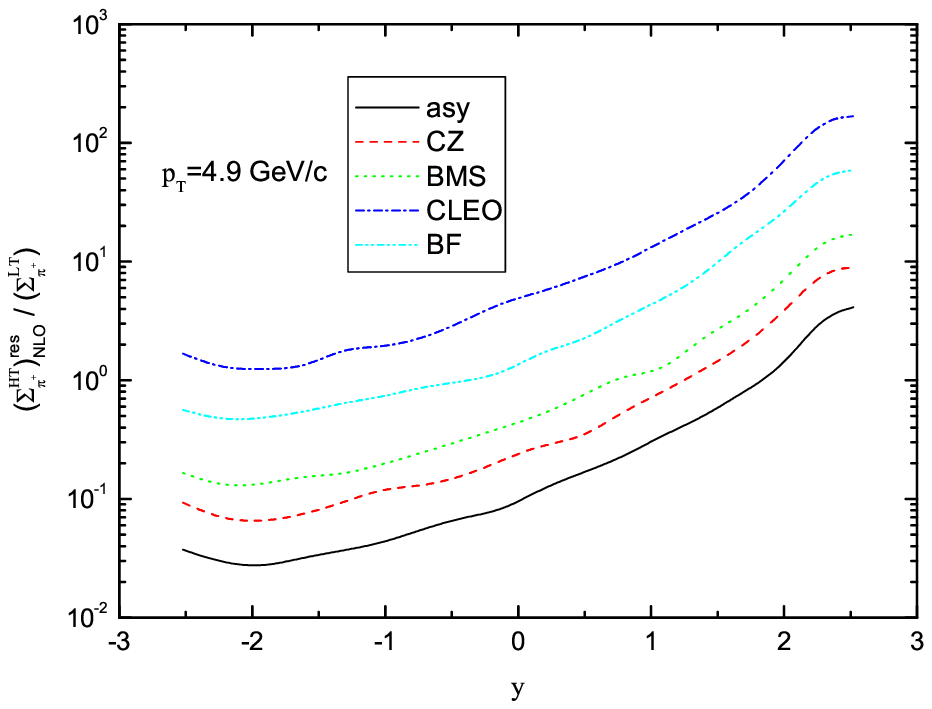}}
\vskip-0.2cm \caption{Ratio
$(\Sigma_{\pi^{+}}^{HT})_{NLO}^{res}/(\Sigma_{\pi^{+}}^{LT})$, as
a function of the $y$ rapidity of the pion at the  transverse momentum
of the pion $p_T=4.9\,\, GeV/c$, at the c.m. energy $\sqrt
s=62.4\,\, GeV$.} \label{Fig8}
\vskip-0.4cm
\vskip 0.8cm \epsfxsize 11.8cm \centerline{\epsfbox{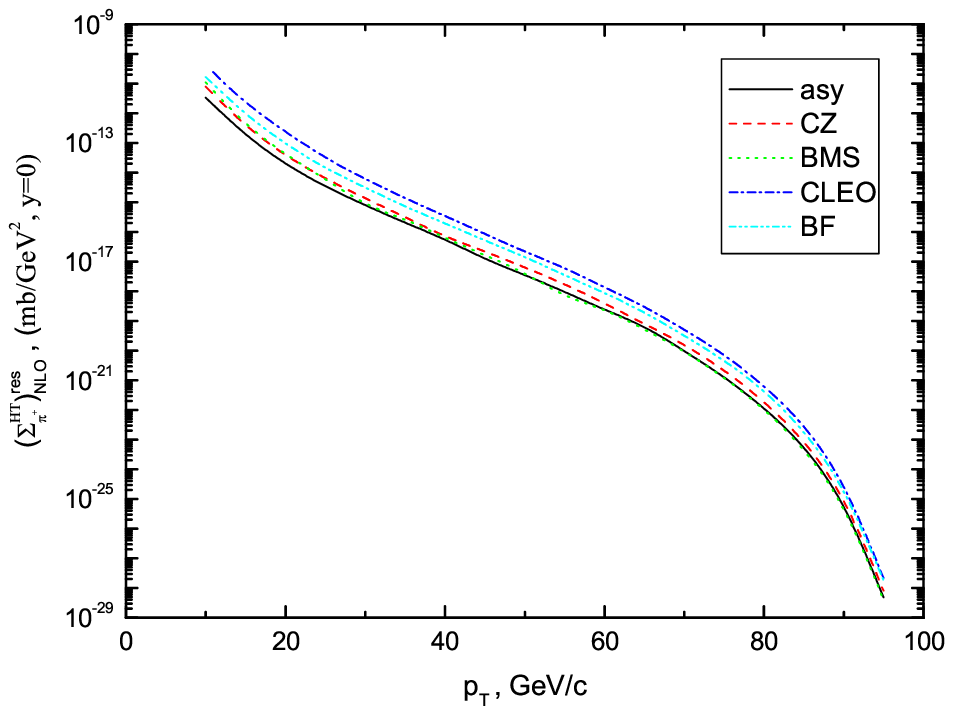}}
\vskip-0.2cm \caption{Next-to-leading order higher-twist $\pi^{+}$ production cross
section $(\Sigma_{\pi^{+}}^{HT})_{NLO}^{res}$ as a function of the $p_{T}$ transverse
momentum of the pion at the c.m.energy  $\sqrt s=200\,\,
GeV$.} \label{Fig9}
\end{figure}

\newpage

\begin{figure}[htb]
\vskip-1.2cm \epsfxsize 11.8cm \centerline{\epsfbox{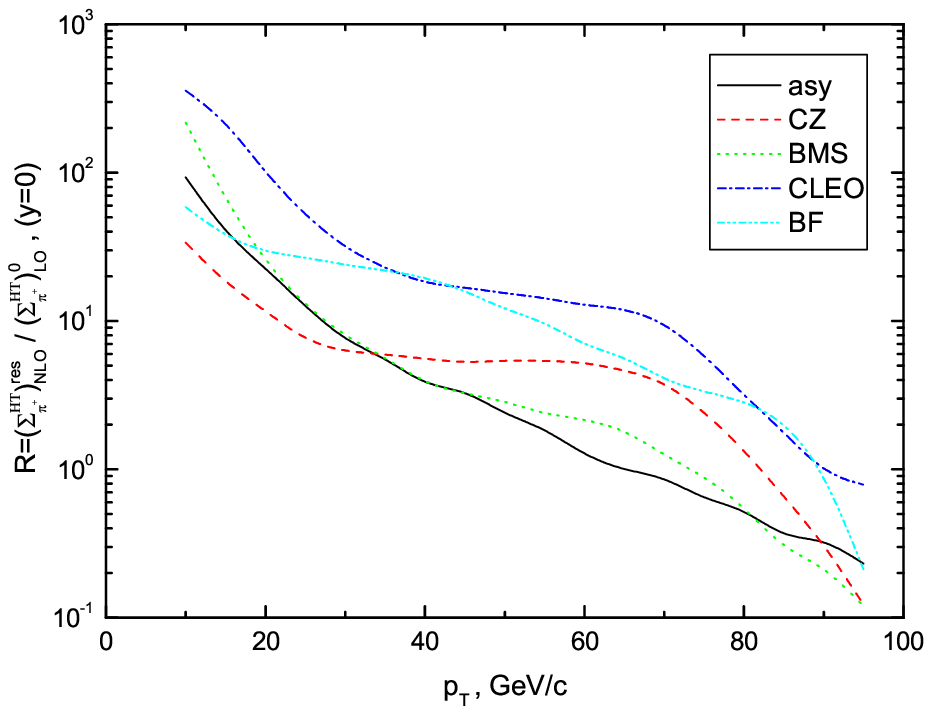}}
\vskip-0.2cm \caption{Ratio
$R=(\Sigma_{\pi^{+}}^{HT})_{NLO}^{res}/(\Sigma_{\pi^{+}}^{HT})_{LO}^{0}$, where
LO and NLO higher-twist contribution are calculated for the pion rapidity $y=0$
at the c.m.energy $\sqrt s=200\,\,GeV$ as a function of the pion
transverse momentum, $p_{T}$.} \label{Fig10} \vskip 1.8cm \epsfxsize 11.8cm
\centerline{\epsfbox{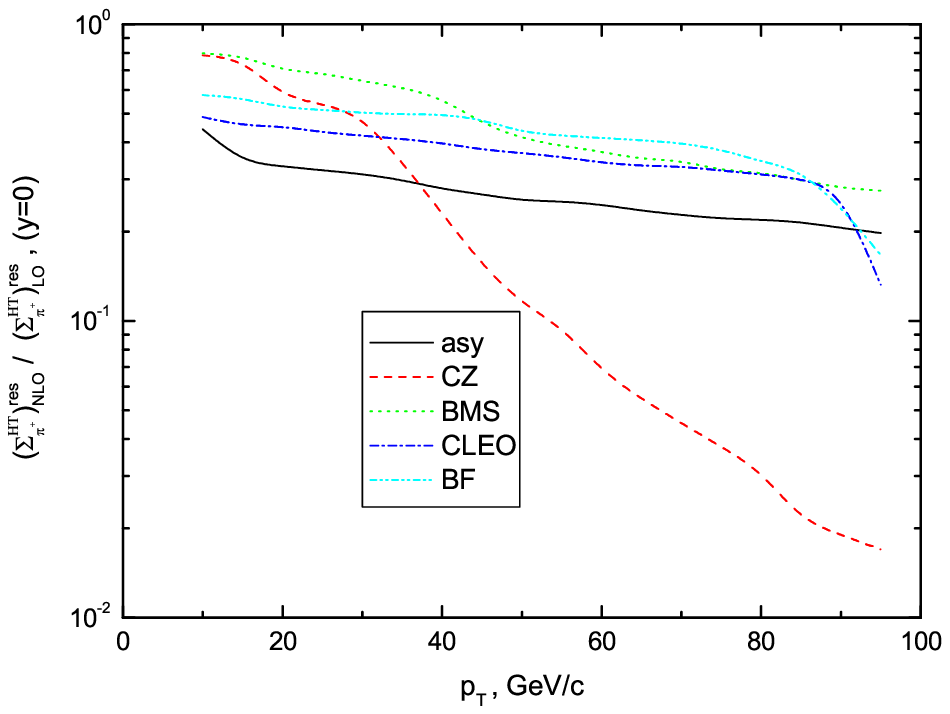}} \vskip-0.05cm \caption{Ratio
$(\Sigma_{\pi^{+}}^{HT})_{NLO}^{res}/(\Sigma_{\pi^{+}}^{HT})_{LO}^{res}$, where
higher-twist contribution are calculated for the pion rapidity $y=0$
at the c.m.energy $\sqrt s=200\,\,GeV$ as a function of the pion
transverse momentum, $p_{T}$.} \label{Fig11}
\end{figure}

\newpage

\begin{figure}[htb]
\vskip-1.2cm\epsfxsize 11.8cm \centerline{\epsfbox{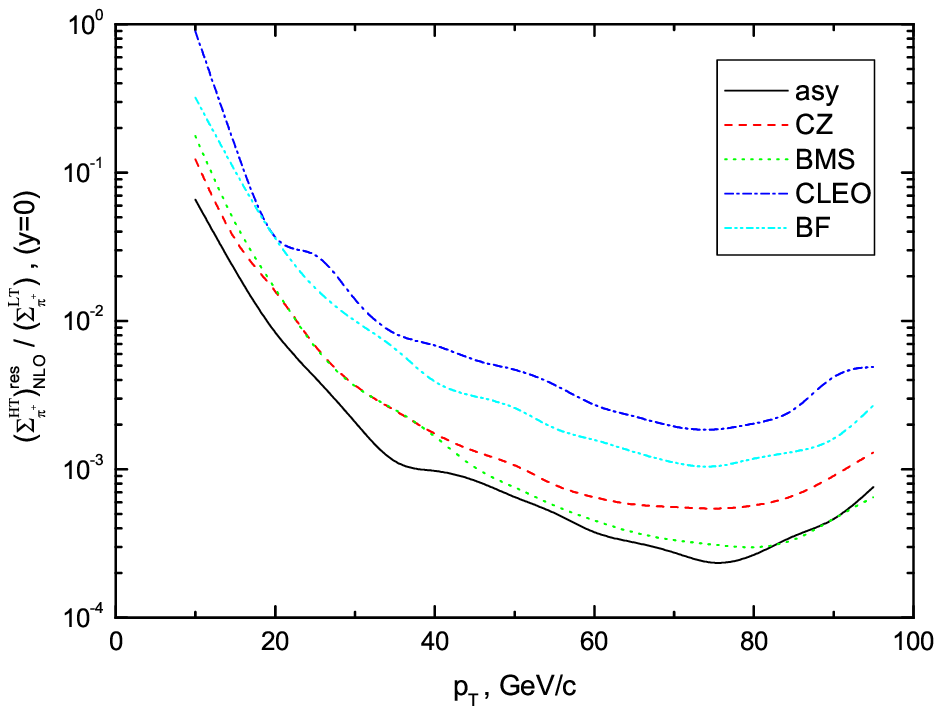}}
\vskip-0.2cm \caption{Ratio
$(\Sigma_{\pi^{+}}^{HT})_{NLO}^{res}/(\Sigma_{\pi^{+}}^{LT})$, as a function of the
$p_{T}$ transverse momentum of the pion
at the c.m.energy $\sqrt s=200\,\,GeV$.}
\label{Fig12} \vskip 1.8cm
\epsfxsize 11.8cm \centerline{\epsfbox{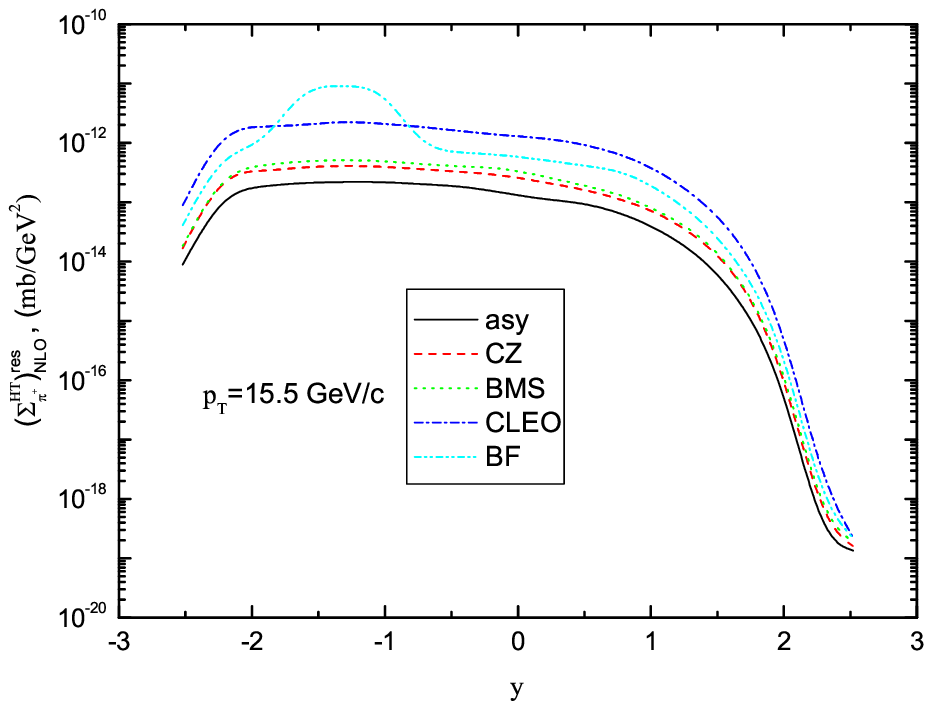}} \vskip-0.05cm
\caption{Next-to-leading order higher-twist $\pi^{+}$ production cross
section $(\Sigma_{\pi^{+}}^{HT})_{NLO}^{res}$, as a function of the $y$
rapidity of the pion at the  transverse momentum of the pion
$p_T=15.5\,\, GeV/c$, at the c.m. energy $\sqrt s=200\,\, GeV$.} \label{Fig13}
\end{figure}

\newpage

\begin{figure}[htb]
\vskip-1.2cm\epsfxsize 11.8cm \centerline{\epsfbox{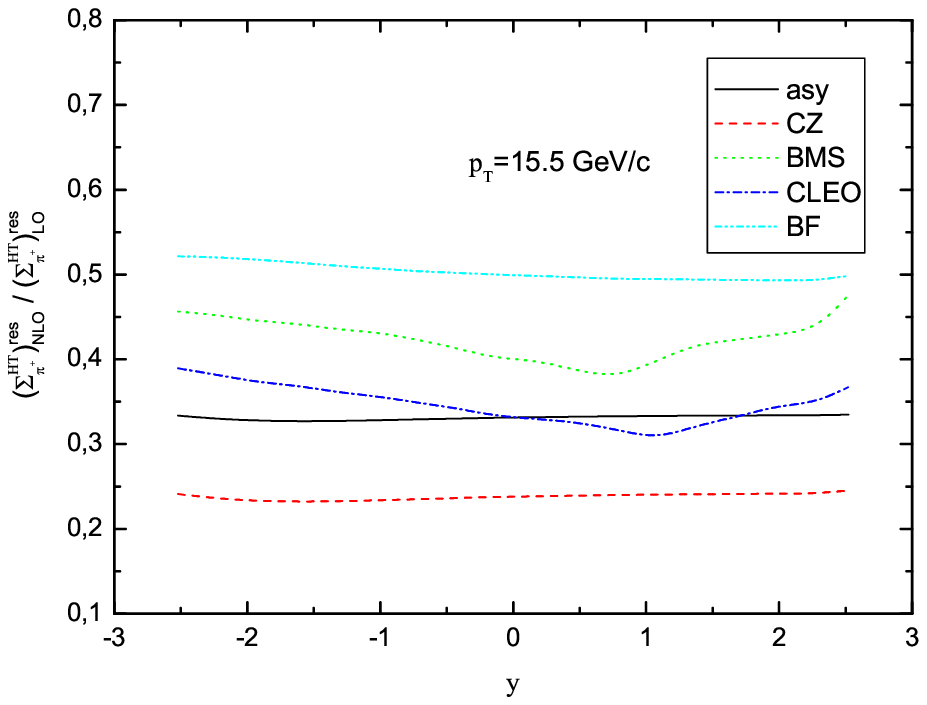}}
\vskip-0.2cm \caption{Ratio
$(\Sigma_{\pi^{+}}^{HT})_{NLO}^{res}/(\Sigma_{\pi^{+}}^{HT})_{LO}^{res}$,
as a function of the $y$ rapidity of the pion at the  transverse momentum
of the pion $p_T=15.5\,\, GeV/c$, at the c.m. energy $\sqrt
s=200\,\, GeV$.} \label{Fig14} \vskip 1.8cm
\epsfxsize 11.8cm \centerline{\epsfbox{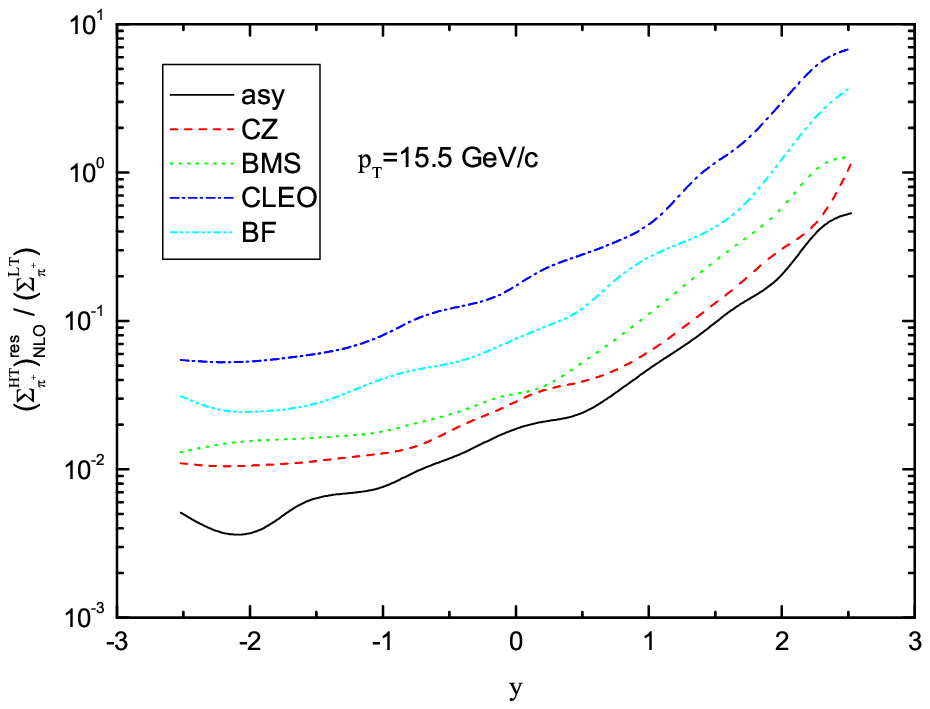}} \vskip-0.05cm
\caption{Ratio
$(\Sigma_{\pi^{+}}^{HT})_{NLO}^{res}/(\Sigma_{\pi^{+}}^{LT})$, as
a function of the $y$ rapidity of the pion at the  transverse momentum
of the pion $p_T=15.5\,\, GeV/c$, at the c.m. energy $\sqrt
s=200\,\, GeV$.} \label{Fig15}
\end{figure}

\end{document}